\documentclass[11pt]{article}
\pdfoutput=1    

\usepackage{setspace}
\usepackage[T1]{fontenc}                
\usepackage{microtype} 				    
\usepackage{anyfontsize}                
\usepackage{amssymb,amsmath,amscd}
\usepackage[UKenglish]{babel}
\usepackage{titlesec}
\usepackage{graphicx,xcolor,xparse}
\usepackage{enumitem}
\usepackage{multirow}
\usepackage[absolute]{textpos}
\usepackage[margin=1.5em,labelfont=bf]{caption}
\usepackage{cite}
\usepackage{mathrsfs}                           
\usepackage[hang,flushmargin]{footmisc}         
\usepackage{orcidlink}                          

\usepackage{hyperref}                           
\hypersetup{
	colorlinks=true,
	linkcolor=dark-blue,
	citecolor=dark-red,
	urlcolor=dark-green,
	linktoc=all
}

\graphicspath{{./figures/}}

\usepackage[vcentermath]{youngtab}
\usepackage{tikz} 
\usepackage{tikz-cd} 
\usetikzlibrary{matrix,calc} 
\usetikzlibrary{decorations.markings,shapes.geometric,shapes.misc} 
\usetikzlibrary{decorations.pathreplacing,positioning} 
\usetikzlibrary{arrows}

\usepackage{geometry} 					
\numberwithin{equation}{section} 

\titleformat{\section}[block]{\Large\bfseries\centering}{\thesection}{1em}{} 
\titleformat{\subsection}[block]{\bfseries}{\thesubsection}{1em}{} 

\titlespacing*{\section}{0pt}{1em}{1em}
\titlespacing*{\subsection}{0pt}{0.75em}{0.75em}

\addto\captionsUKenglish{
  \renewcommand{\contentsname}%
    {Table of Contents}%
}

\definecolor{dark-gray}{gray}{0.20}
\definecolor{gray}{gray}{0.30}
\definecolor{light-gray}{gray}{0.80}
\definecolor{dark-red}{rgb}{0.7,0,0}
\definecolor{dark-green}{rgb}{0.1,0.4,0}
\definecolor{dark-blue}{rgb}{0.3,0.3,0.7}
\definecolor{light-blue}{rgb}{0.8,0.8,1}
\definecolor{cardinal}{rgb}{0.6,0,0}
\definecolor{darkgreen}{rgb}{0,0.5,0}
\definecolor{golden}{rgb}{0.92, 0.7, 0}
\definecolor{midnight}{rgb}{0, 0, 0.5}
\definecolor{darkblue}{rgb}{0.2, 0, 0.8}
\definecolor{forestgreen}{rgb}{0.13, 0.55, 0.13}
\definecolor{darkred}{rgb}{0.7,0,0}

\def\mop#1{\mathop{\rm #1}\nolimits}

\def\diag{\mop{diag}}
\def\ii{{\rm i}}

\def\AdS{{\mop{AdS}}}

\def\rank{\mop{rank}}
\newcommand{\ds}{{\rm d}s}
\newcommand{\dd}{{\rm d}}
\newcommand{\DD}{{\rm D}}
\newcommand{\e}{\mathrm{e}}

\newcommand\Tr{\mathrm{Tr}\,}

\newcommand{\dvol}{{\rm vol}}
\newcommand{\vol}{\mathbf{V}}

\newcommand{\disc}{\mathbb{D}}

\newcommand{\f}[2]{\frac{#1}{#2}}

\newcommand{\vev}[1]{\left\langle {#1} \right\rangle}

\newcommand{\nn}{\nonumber}
\newcommand{\wti}[1]{\widetilde{#1}}

\def\overleftrightarrow#1{\vbox{\ialign{##\crcr
	$\leftrightarrow$\crcr\noalign{\kern-0pt\nointerlineskip}
	$\hfil\displaystyle{#1}\hfil$\crcr}}}
\def\slashed#1{{\ooalign{\hfil\hfil/\hfil\cr $#1$}}}

\newcommand\bC{\mathbf{C}}

\newcommand\bH{\mathbf{H}}

\newcommand\bP{\mathbf{P}}

\newcommand\bR{\mathbf{R}}

\newcommand\bZ{\mathbf{Z}}

\newcommand\cA{\mathcal{A}}

\newcommand\cC{\mathcal{C}}

\newcommand\cF{\mathcal{F}}

\newcommand\cH{\mathcal{H}}
\newcommand\cI{\mathcal{I}}

\newcommand\cL{\mathcal{L}}
\newcommand\cM{\mathcal{M}}
\newcommand\cN{\mathcal{N}}
\newcommand\cO{\mathcal{O}}
\newcommand\cP{\mathcal{P}}

\newcommand\cS{\mathcal{S}}

\newcommand\fg{\mathfrak{g}}

\newcommand\ft{\mathfrak{t}}

\newcommand\SO{\mathrm{SO}}

\newcommand\UU{\mathrm{U}}
\newcommand\SU{\mathrm{SU}}

\renewcommand\sl{\mathfrak{sl}}
\newcommand\so{\mathfrak{so}}

\newcommand\psl{\mathfrak{psl}}
\newcommand\gl{\mathfrak{gl}}

\newcommand{\TODO}[2]{
   \ifx&#1&%
       {
        \textcolor{darkblue}{\tt {\bf TODO:} #2}
        } 
   \else
        {
       \textcolor{darkblue}{\tt {\bf TODO@{#1}:} #2}
        } 
\fi
}

\title{~\vspace{10mm}\\
\fontsize{20pt}{23pt}\selectfont\textbf{Holographic Generalised Gukov-Witten Defects}\vspace{10mm}}

\author{\Large{Pieter Bomans\textsuperscript{\orcidlink{0000-0002-0907-9830}  $\bullet$} \hspace{2mm} and \hspace{2mm} Lorenzo Tranchedone\textsuperscript{\orcidlink{0009-0004-8199-6539}  $\circ$}}\\[10mm]
    \large $~^\bullet$Mathematical Institute, University of Oxford\\[1mm]
    \normalsize Andrew Wiles Building, Radcliffe Observatory Quarter\\
    \normalsize Woodstock Road, Oxford, OX2 6GG, U.K.\\[5mm]
    \large $~^\circ$Rudolf Peierls Centre for Theoretical Physics, University of Oxford\\[1mm]
    \normalsize Beecroft Building, Clarendon Laboratory,\\
    \normalsize Parks Road, University of Oxford, OX1 3PU, U.K.\\[5mm]
    \texttt{\normalsize\href{mailto:pieter.bomans@maths.ox.ac.uk}{pieter.bomans@maths.ox.ac.uk} \hspace{2em}\href{mailto:lorenzo.tranchedone@physics.ox.ac.uk}{lorenzo.tranchedone@physics.ox.ac.uk}}\\
}

\date{}

\begin{document}

\pagestyle{empty}

\maketitle
\thispagestyle{empty}

\vspace{\stretch{1}}

\begin{abstract}
\noindent Spindle and disc solutions have received significant attention in recent literature. However, it is not entirely clear what these supergravity solutions represent in the dual SCFT. In this work we elucidate this issue by considering a different global completion of the spindle solution, focusing on a single conical defect. We argue that these solutions can be interpreted as the insertion of a single (generalised) Gukov-Witten defect in the bulk SCFT. We give compelling evidence for this claim by uplifting the solutions to type IIB supergravity and showing that the five-dimensional $\f12$-BPS solutions uplifts to the bubbling surface operators solutions. Their less supersymmetric cousins give rise to a large class of holographic generalised Gukov-Witten surface operators for which we perform a detailed geometric analysis. To corroborate these claims we compute a variety of holographic observables such as the renormalised action, stress tensor one-point function and defect entanglement entropy. In addition, we extend the holographic inflow mechanism to account for bulk-to-defect inflow to study the symmetries and anomalies of the bulk and defect.  
\end{abstract}

\vspace{\stretch{3}}
 
\newpage

{ 
\hypersetup{linkcolor=black}
\setcounter{tocdepth}{2}
\tableofcontents
}

\setcounter{page}{0}

\clearpage
\pagestyle{plain}

\section{Introduction}      %
\label{sec:intro}                       %

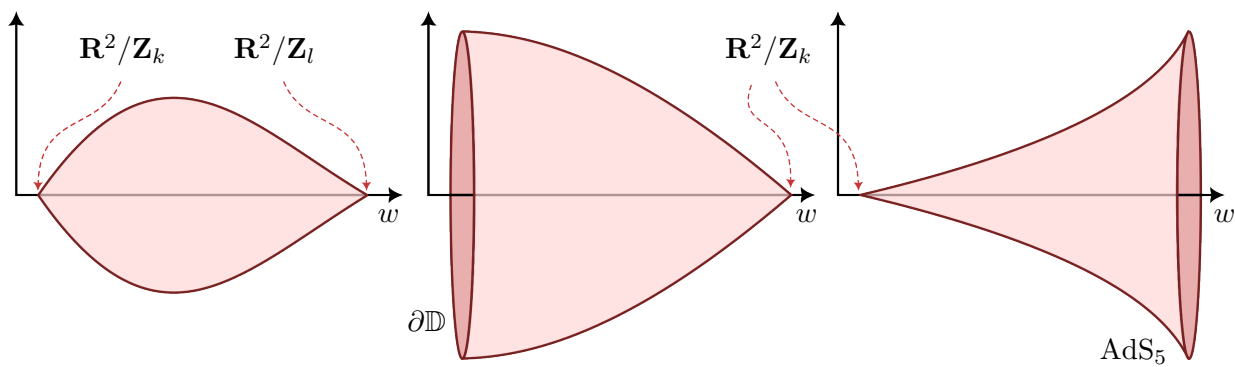
\begin{figure}[!htb]
    \centering
    \begin{tikzpicture}
        \node {\includegraphics[width=\linewidth]{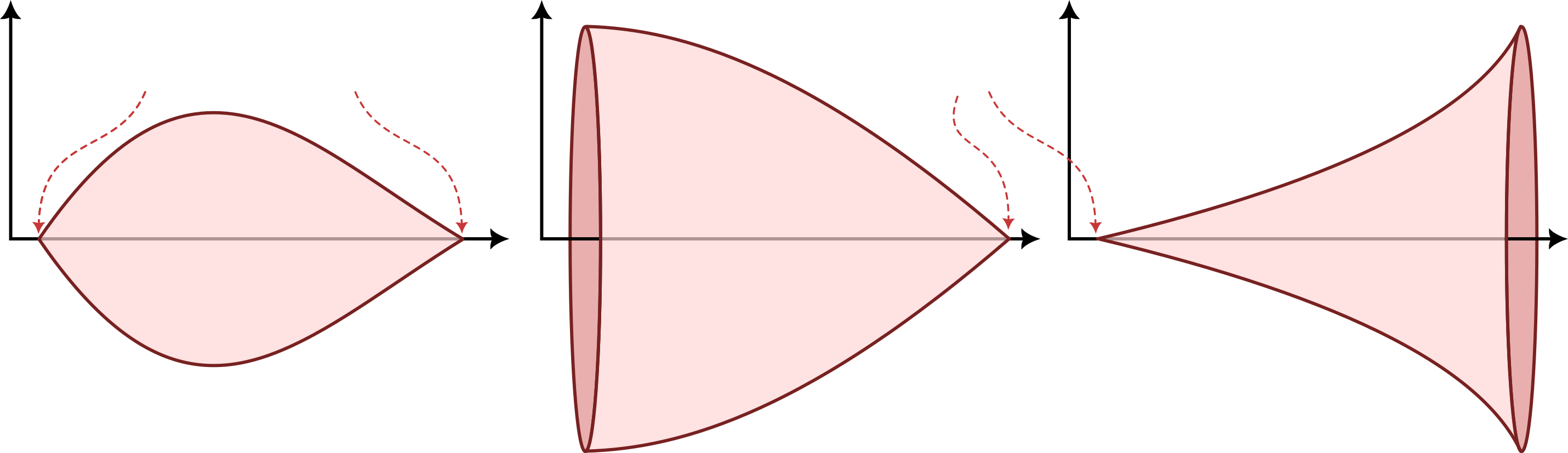}};
        \node at (-6.5,1.8) {$\bR^2/\bZ_{k}$};
        \node at (2,1.8) {$\bR^2/\bZ_{k}$};
        \node at (-4.5,1.8) {$\bR^2/\bZ_{l}$};
        \node at (-2.5,-1.8) {$\partial\disc$};
        \node at (6.8,-2.2) {$\AdS_5$};
        \node at (-3,-0.4) {$w$};
        \node at (2.5,-0.4) {$w$};
        \node at (8,-0.4) {$w$};
    \end{tikzpicture}
    \caption{A graphical representation of the spindle, disc and (non-compact) defect (from left to right). The spindle is a sphere with two conical defects, while the disc has a conical defect and a boundary. Finally, the defect solution has a conical defect but unlike the two previous ones approaches $\AdS_5$ in the asymptotic large $w$ region.}
    \label{fig:SpindleDiscDefect}
\end{figure}

Spindle and disc solutions have received significant attention in recent literature. See \cite{Ferrero:2020laf,Ferrero:2020twa,Hosseini:2021fge,Boido:2021szx,Bah:2021mzw,Bah:2021hei,Ferrero:2021wvk,Cassani:2021dwa,Couzens:2021tnv,Suh:2021ifj,Suh:2021aik,Ferrero:2021ovq,Couzens:2021rlk,Faedo:2021nub,Ferrero:2021etw,Couzens:2021cpk,Bah:2022yjf,Couzens:2022yjl,Amariti:2023mpg,Hristov:2023rel,Bomans:2023ouw,Ferrero:2024vmz,Boisvert:2024jrl,Bomans:2024mrf} for a necessarily incomplete set of references. These solutions present a generalisation of the usual lore of lower dimensional supergravity by allowing for certain (mild) singularities. In particular, they include conical defects. Unlike in \cite{Bobev:2019ore} where similar conical defects were introduced on higher genus Riemann surfaces, the spindle and disc represent respectively a sphere with (one or) two conical defects at the poles, or a topological disc with one defect at the centre. See Figure \ref{fig:SpindleDiscDefect} for a graphical representation. Provided that the orders of the conical defects are co-prime, these are so-called bad orbifold \cite{thurston2022geometry}, which do not admit a constant curvature metric. For this reason, such solutions preserve supersymmetry in a novel way. In fact for the spindle solutions there are often two options, dubbed the twist and anti-twist \cite{Ferrero:2021etw}.

By now, this class of supergravity solutions has achieved widespread acceptance in the supergravity community. However, it remains unclear how to interpret them holographically. Nonetheless, in particular cases progress in this direction has been made. The seven-dimensional disc solution has been understood as the holographic dual to a class of Argyres-Douglas SCFTs \cite{Bah:2021mzw,Bah:2021hei,Bah:2022yjf,Couzens:2022yjl,Couzens:2023kyf}. In these cases, the conical defect can be understood as the insertion of an $\cN=2$ preserving co-dimension two disorder defect, colloquially known as a regular puncture, while the boundary of the disc is dual to an $\cN=2$ preserving co-dimension two defect corresponding to an irregular puncture. Similarly, in \cite{Bomans:2024mrf} it was demonstrated that the conical defects of the seven-dimensional spindle solution can be interpreted as insertions of $\cN=1$ preserving punctures. 

The main claim of this work, in line with the claims above, is that such an interpretation is valid more generally. In particular, we will argue that the conical defects in the five-dimensional spindle and disc solutions correspond to a class of generalised Gukov-Witten defects \cite{Gukov:2006jk,Gukov:2008sn}. To argue for these claims we will zoom in to a single conical defect of the spindle and study a class of holographic defect solutions. Similar defect solutions have been considered in gauged supergravity in \cite{Gutperle:2019dqf,Gutperle:2023yrd,Gutperle:2022pgw,Capuozzo:2023fll,Arav:2024exg,Arav:2024wyg} while in ten- or eleven-dimensional supergravity several classes of $\f12$-BPS defects have been discussed as bubbling defect solutions \cite{Gomis:2006sb,Yamaguchi:2006te,Gomis:2007fi,Lunin:2007ab,Drukker:2008wr,Drukker:2008jm,Lin:2004nb,Lunin:2015hma,Choi:2024ktc}. In this work we generalise the analysis of said papers by starting from a class of gauged supergravity solutions with a conical defect in the bulk. Unlike \cite{Gutperle:2019dqf,Arav:2024exg} this generalisation allows us to holographically access a class of $\f12$-BPS defects and significantly enlarge the class of accessible, less supersymmetric, defects.

\paragraph{Summary of this work}~\\

In this work we focus on surface defect operators in four-dimensional $\cN=4$ super Yang Mills (SYM) theory. The dual supergravity solutions are therefore constructed in five-dimensional maximal gauged supergravity. We start from the basic (local) spindle solution, constructed in \cite{Ferrero:2020laf,Boido:2021szx} which can be constructed in the STU model and subsequently generalise these solutions by including three additional complex scalars which partly break the $\UU(1)^3$ global symmetry. Note that similar defect solutions exist in half-maximal and minimal gauged supergravity hence our work allows for immediate generalisations to large classes of four-dimensional $\cN=2$ and $\cN=1$ SCFTs with defect insertions.

Unlike for the spindle or disc discussed above, we consider a non-compact global completion with an asymptotic $\AdS_5$ region which can be interpreted as holographic surface defects. These solutions are very similar in spirit to the solutions considered in \cite{Gutperle:2019dqf,Arav:2024exg} with the important difference that we allow for conical defects in the five-dimensional bulk while insisting on a smooth boundary. This relaxed smoothness condition allows us to go beyond the class of surfaces constructed in \cite{Gutperle:2019dqf,Arav:2024exg} and most importantly includes a class of $\f12$-BPS defects which are ruled out when insisting on having a smooth bulk.

Uplifting our solutions to ten-dimensional type IIB supergravity, this class of $\f12$-BPS surface defect solutions can be embedded in the bubbling surface ansatz of \cite{Gomis:2007fi,Drukker:2008wr,Drukker:2008jm,Lin:2004nb}. Doing so we can precisely identify the type of surface defect that our solutions correspond to. As expected, the five-dimensional supergravity does not include enough degrees of freedom to encode any type of defect. Instead, we find that the solutions constructed in the STU model describe axially symmetric defects labelled by rectangular Young Tableaux. Including the additional scalars breaks the axial symmetry and therefore allows us to describes a larger class of surface defects. 

Reducing the supersymmetry preserved by the surface defects, our supergravity solutions give rise to a large class of novel $\f14$- and $\f18$-BPS surface defects. These surface defects are much less studied in the field theory literature and consequently results are sparse. Our supergravity solutions therefore provide a concrete road to explore such defects. While the conical defects in the $\f12$-BPS case uplift to $\bR^4/\Gamma$ singularities, with $\Gamma$ a discrete subgroup of $\SU(2)$, their less supersymmetric cousins uplift to $\bR^6/\Gamma$ singularities, with $\Gamma$ a discrete subgroup of $\SU(3)$ or $\UU(3)$ respectively for $\f14$- and $\f18$-BPS surface defects. By carefully studying the orbibundle data of the lower-dimensional supergravity solutions as well as their uplifts, we can describe a large class of generalised holographic Gukov-Witten defects.

To provide further evidence for the conjectured identification of our supergravity solutions with surface defects, we compute a variety of holographic observables including the renormalised on-shell action, the stress tensor one-point function and the defect entanglement entropy. These quantities in turn allow us to holographically describe two defect anomalies in terms of the supergravity parameters $w_0$ and $q_I$,
\begin{equation}
    b = 3N^2 (1-w_0) \,,\qquad \quad d_2 = -3N^2 (q_1+q_2+q_3)\,,
\end{equation}
which appear as follows in the defect Weyl anomaly \cite{Jensen:2013lxa,Jensen:2019ano,Chalabi:2020iie,Chalabi:2021jud,Chalabi:2022wey},
\begin{equation}
    T^\mu_\mu\bigg|_{\Sigma} = -\f1{24\pi}\left( b R_\Sigma + d_1 \Pi^2 - d_2 W^{ab}{}_{ab} \right) + \cdots 
\end{equation}
Where $R_\Sigma$ is the Ricci curvature of the defect world-volume, $\Pi^2$ is a particular contraction of the extrinsic curvature and $W^{ab}{}_{ab}$ is the contraction of the bulk Weyl tensor, pulled back to the defect world volume. In this formula $b$ is the only type A anomaly. For the $\f12$-BPS defects, the expressions above agree precisely with the expected defect anomalies for (axially symmetric) Gukov-Witten defects \cite{Gomis:2007fi,Drukker:2008wr,Drukker:2008jm,Choi:2024ktc} while for their less supersymmetric cousins these provide explicit predictions from holography.

In addition to matching the anomalies above, we extend the holographic inflow formalism developed in \cite{Bah:2018gwc,Bah:2019rgq,Bah:2019jts,Bah:2020jas} to bulk-to-defect inflow. In contrast to these works, our solutions contains both an $\AdS_3$ region corresponding to the defect and an asymptotic $\AdS_5$ region corresponding to the bulk theory. However, we can deal with this in exactly the same way as in the papers above, by excising a disc from the (non-compact) internal space corresponding to the asymptotic region as well as a disc surrounding the singularity corresponding to the defect. The remaining cylindrical region does not contribute to the anomaly polynomial so this leaves us with two contributions, a six-form anomaly polynomial corresponding to the bulk theory coming from the asymptotic region as well as a four-form defect anomaly polynomial coming from the singular disc. Both are obtained by integrating the auxiliary eleven form $\cI_{11}$ over the relevant part of the internal space \cite{Bah:2020jas}. Doing so we reproduce the defect central charge $b$ in an alternative way, confirming the result above. In addition, by carefully analysing the blow-up of the singularities arising in the bulk we can determine the flavour symmetries localised to the defect. In particular, we show that for the $\f12$-BPS defects this precisely matches with the field theory expectation, reviewed in Section \ref{sec:QFT-defects}.

\paragraph{Outlook}~\\

Our work makes progress on the outstanding question to identify the field theory interpretation of spindle solutions. This paper identifies the conical defects appearing in the spindle geometries as generalised Gukov-Witten defects. Similarly, in \cite{Bah:2021mzw,Bah:2021hei,Bah:2022yjf,Couzens:2022yjl,Couzens:2023kyf,Bomans:2024mrf}, defects in the seven-dimensional spindle and disc solution were interpreted as co-dimension two disorder defects in the $\cN=(2,0)$ theory. A natural conjecture extending our work is that spindle and disc solutions in any dimension should be interpreted as adding co-dimension two defects to the bulk theory. In the four-dimensional gravitation setup, dual to the three-dimensional ABJM theory we make progress in this direction by identifying the conical defects in the spindle solution as vortex lines \cite{Boido:coming}. See also \cite{Arav:2024wyg} for a holographic discussion of a class of closely related monodromy defects in ABJM. A similar interpretation should exist in the six-dimensional spindle solutions dual to the Seiberg SCFTs. These defects are not well studied in the literature making it an interesting objective for further research.

One notable difference with the setup discussed here and the five-dimensional spindle solutions is that in the latter setup the uplifted geometry is completely regular. This difference arises from our choice to have no conical defects at the boundary, which fixes the $z$-coordinate to be $2\pi$-periodic. Related to this, we expect that if a full holographic RG flow solution, connecting the spindle solution with an asymptotic $\AdS_5$ region, would be known, this solution would contain conical singularities on the $\bR^{1,1}\times S^2$ boundary. Forcing these singularities to be absent we expect to retrieve the conical singularities in the ten-dimensional bulk. 

This difference is reminiscent of the story proposed in \cite{Jeong:2020uxz}, where two alternative prescriptions to describe surface defects were proposed. On the one hand, surface defects are described by specifying the singularities along a co-dimension two locus, similar as the disorder defects discussed in this work, while on the other hand they show that there is an equivalent description of these defects by including conical defects in space-time. It would be interesting to understand the connection of this work with our setup to make the above story more precise. A starting point to understand this setup would be to study the spindle index discussed in \cite{Inglese:2023wky,Colombo:2024mts,Mauch:2024uyt}. In these works ABJM is localised on $S^1 \times \Sigma$, giving rise to an index that closely resembles the standard index on $S^1\times S^2$ with the insertion of vortex lines at the poles of the $S^2$.

In this work we constructed five-dimensional supergravity solutions and subsequently uplifted them to ten dimensions. In the $\f12$-BPS case, we found that our solutions can be embedded into the bubbling surface defect setup and give rise to a subset of possible Gukov-Witten defects. The ten-dimensional setup is richer and allows for generalisation to holographically describe any Gukov-Witten surface defect in $\cN=4$. It would be interesting to study the ten-dimensional setup corresponding to the $\f14$-BPS defects and embed them in a generalised bubbling ansatz derived in \cite{Jia:2023iaa}. The PDEs arising in this setup are much more involved and a similar classification of boundary conditions is currently beyond reach. However, starting from the seed solutions provided by our five-dimensional setup could provide a concrete point to generalise our solutions and obtain holographic control over a extended class of surface operators.

Finally, in this work we discussed defect solutions with a conical defect in the bulk and conjectured these to be dual to a class of Gukov-Witten defects. Focussing on the $\f12$-BPS case, this class of defects are known as tame or regular defects, since they correspond to tame ramifications in an associated Hitchin system \cite{Gukov:2006jk,Gukov:2008sn}. Similar to the class $\cS$ story, there is a whole set of irregular or wild punctures, corresponding to wild ramifications \cite{Witten:2007td}. These defects are not well-studied hence it would be very interesting to get a holographic grip on them. A natural guess would be that they correspond to $\f12$-BPS disc-defect solutions with a disc-like boundary instead of a conical defect in the bulk. However, a preliminary analysis suggests that these solutions behave rather different than expected. We leave this as an open question for future explorations.  

\paragraph{Structure of this paper}~\\

The remainder of this paper is organised as follows. We start in Section \ref{sec:QFT-defects} with a review of generalised Gukov-Witten defects in $\cN=4$ SYM and define various defect observables. In Section \ref{sec:solutions}, we continue by constructing the relevant supergravity solutions. To facilitate the analysis of these solutions we uplift them to type IIB supergravity in Section \ref{sec:uplift} and demonstrate that the $\f12$-BPS defect solutions uplift to a class of bubbling surface operator solutions, while their less supersymmetric cousins provide a holographic description of a class of generalised Gukov-Witten surface operators. To further corroborate our claims, we proceed to compute a variety of holographic defect observables in Section \ref{sec:observables} where we introduce the framework of holographic bulk-to-defect inflow. Various technical details are deferred to Appendices \ref{app:EOM-BPS}-\ref{app:uplift}.

\textit{Note:} Part of this paper is based on results obtained in the master thesis of the second author \cite{ThesisLT}.

\section{Surface operators in \texorpdfstring{$\cN=4$}{N=4} SYM}
\label{sec:QFT-defects}                    

One can divide extended operators in two groups, order operators or disorder operators. Order operators are defined by a field theory supported on a codimension $p$ surface $\Sigma\subset \bR^d$ coupled to the $d$-dimensional ambient theory. Disorder defects on the other hand are characterised by a codimension $p$ singularity of the $d$-dimensional fields along $\Sigma$.\footnote{In Lagrangian field theories this division is analogous to the division of local operators in order operators defined as electric fields or disorder operators such as monopoles, defined through their singular behaviour at a point.} In this work we will focus on disorder type surface operators in $\SU(N)$ $\cN=4$ SYM theory. 

The $\cN=4$ superconformal algebra in four dimensions is given by $\psl(4|4)\supset \sl(4)\times \sl(4)_R$. Note that we specify the complexified superalgebra and maximal bosonic subalgebra. Inserting a conformal surface operator breaks the bulk conformal symmetry to  $\sl(4)~\rightarrow~\sl(2)~\times~\sl(2)~\times~\gl(1)_C$, where $\sl(2)\times \sl(2)$ is the two-dimensional conformal algebra while $\gl(1)_C$ is the diagonal combination of the an internal symmetry and a rotation transverse to the defect. The defects considered in this work preserve $\f12$, $\f14$ or $\f18$ of the supercharges which organise into the following superalgebras, 
\begin{itemize}
    \item $\f12$-BPS: (small) $\cN=(4,4)$ superconformal algebra $\psl(2|2)\times\psl(2|2) \ltimes \gl(1)_C$.
    \item $\f14$-BPS: $\cN=(2,2)$ superconformal algebra $\sl(2|1)\times\sl(2|1) \ltimes \gl(1)_C$.
    \item $\f18$-BPS: $\cN=(0,2)$ superconformal algebra $\sl(2|1)\times\sl(2) \ltimes \gl(1)_C\times \gl(1)_F$.
\end{itemize}
By definition conformal surface operators preserve the respectively two-dimensional superconformal group, however as we will see below, the transverse rotations $\gl(1)_C$ as well as the additional flavour symmetry $\gl(1)_F$ can be broken. These types of defects were first introduced in the seminal works \cite{Gukov:2006jk,Gukov:2008sn} where the $\f12$-BPS defects were studied in detail. Holographically it will be more convenient to perform a conformal transformation and consider $\cN=4$ SYM on $\AdS_3 \times S^1$ and insert the defect at the boundary of $\AdS_3$. From now on we will restrict to this setup. 

The field content of $\cN=4$ SYM is given by a gauge field $A$ and six scalar fields $X_i$ transforming in the vector representation of $\so(6)_\bC \simeq \sl(4)_R$. In addition, the theory contains four fermions $\lambda_a$ transforming in the $\mathbf{4}$ of $\sl(4)$. All fields transform in the adjoint representation of the gauge group. Given the symmetry breaking patterns discussed above, it will be useful to organise the scalars into three complex scalars $\Phi_I =  \f{1}{\sqrt 2}(X_{2I-1}+ \ii X_{2I})$.

As explained in \cite{Gukov:2006jk}, in the case of the $\f12$-BPS defects two of the complex scalars necessarily vanish and the BPS equations relevant for our setup reduce to Hitchin's equations. For the less supersymmetric defects respectively one or two additional scalars can be non-vanishing. In this case the BPS equations reduce to a generalisation of Hitchin's equations analogous to the one in \cite{Xie:2013gma}. Surface operators can then be described as singular boundary conditions for these (generalised) Hitchin equations. While the most general boundary conditions have not been classified, a subset of these less supersymmetric defects is obtained by applying the same type of boundary conditions to the additional scalars as for the $\f12$-BPS case, described in \cite{Gukov:2006jk,Gomis:2007fi}.


A disorder surface operator $V_\Sigma$ is characterised by a vortex-like singularity for the gauge field near $\Sigma$. Near this surface, the gauge field configuration breaks the gauge group $G$ to a Levi subgroup $L = {\rm S}\left(\prod \UU(N_l)\right)\subset \SU(N)$.\footnote{In fact, surface operators are labelled by conjugacy classes, since gauge transformations act by conjugation $A\to g A g^{-1}$.}
 The singular gauge field configuration near the surface can be written as
\begin{equation}
    A = \begin{pmatrix}
        \alpha_1 \otimes 1_{N_1} & 0 & \cdots & 0\\
        0 & \alpha_2 \otimes 1_{N_2} & \cdots & 0\\
        \vdots  & \vdots & \ddots & \vdots \\
        0 & 0 & \cdots & \alpha_M \otimes 1_{N_M}
    \end{pmatrix}
    \dd\psi\,,
\end{equation}
where $\psi$ is the rotation along the non-contractible $S^1$ and $1_{N_l}$ is the $N_l$ dimensional unit matrix. Since we have $\sum_l N_l = N$, different surface operators can be classified by Young tableaux with $M$ rows of length $N_l$. In addition, the surface operator introduces two-dimensional $\theta$-angles,
\begin{equation}
    \eta = \begin{pmatrix}
        \eta_1 \otimes 1_{N_1} & 0 & \cdots & 0\\
        0 & \eta_2 \otimes 1_{N_2} & \cdots & 0\\
        \vdots  & \vdots & \ddots & \vdots \\
        0 & 0 & \cdots & \eta_M \otimes 1_{N_M}
    \end{pmatrix}\,,
\end{equation}
corresponding to an insertion of the operator $\exp \left(\ii \sum_l \eta_l \int_\Sigma \Tr F_l\right)$ in the path integral. Finally, a $\f12$-BPS surface operator has to preserve $\sl(2)\times \sl(2)$ R-symmetry so in this case one is restricted to turn on a non-trivial field configuration for $\Phi_1$ only
\begin{equation}\label{eq:scalform}
    \Phi_1 = \e^{-\ii\psi}\begin{pmatrix}
        (\beta_1+\ii \gamma_1) \otimes 1_{N_1} & 0 & \cdots & 0\\
        0 & (\beta_2+\ii \gamma_2) \otimes 1_{N_2} & \cdots & 0\\
        \vdots  & \vdots & \ddots & \vdots \\
        0 & 0 & \cdots & (\beta_M+\ii \gamma_M) \otimes 1_{N_M}
    \end{pmatrix}\,.
\end{equation}
Note that turning on a non-trivial profile for the scalar fields breaks the transverse rotation and internal symmetry parameterised by $\gl(1)_C$. As discussed above, when considering $\f14$-BPS or $\f18$-BPS defects we are free to turn on non-trivial profiles for one or two of the remaining complex scalar fields of exactly the same form as in \eqref{eq:scalform}. The surface operators in this class are therefore described by a total of $4M$, $6M$ or $8M$ parameters respectively (up to the action of $S_M$),
\begin{equation}
    \left(\alpha_i,\eta_i,\beta^{(1)}_i, \gamma^{(1)}_i,\beta^{(2)}_i, \gamma^{(2)}_i,\beta^{(3)}_i, \gamma^{(3)}_i \right) \in \left( \ft, {}^L\ft, \ft ,\ft , \ft,\ft\right)\,,
\end{equation}
where $\ft$ denotes the maximal torus of the gauge algebra and ${}^L \ft$ the maximal torus of the Langlands dual of the gauge algebra. 

The $\f12$-BPS punctures introduced above, i.e. $\beta^{(2,3)}=\gamma^{(2,3)}=0$ carry localised global symmetries given by the commutant $H\subset G$ of $L$. In terms of the associated Young tableaux we can write the global symmetry associated to a particular surface operator as 
\begin{equation}\label{eq:localised-flavour}
    H = {\rm S}\prod_{k} \UU(h_k)\,,
\end{equation}
where $h_k$ is the number of columns of height $k$.\footnote{For example, the maximal Levi subgroup $L=G$ is associated with the (almost) trivial surface operator, in this case the commutant only is the centre $Z(G)$. These surfaces represent the topological surfaces generating the $\bZ_N$ one-form symmetry of the $\SU(N)$ $\cN=4$ SYM theory. On the opposite side of the spectrum we have the minimal Levi subgroup $L = T$, where $T$ is the maximal torus inside $G$. This surface operator has an associated $\SU(N)$ global symmetry.} As explained in \cite{Gukov:2006jk,Gukov:2008sn} the classification of surface operators follows from an underlying Hitchin system. The surface operators introduced so far correspond to tame ramifications of the associated Hitchin system. Similar to the classification of punctures in class $\cS$, there should exist more general surface operators corresponding to wild ramifications \cite{Witten:2007td}. These operators should allow for more singular behaviour of the scalars. However, very little is known about such surface operators. It would be interesting to learn more about them, perhaps by generalising the holographic methods developed in this work.

\subsection{Defect observables}

The insertion of a surface operator breaks the four-dimensional superconformal algebra to a sub-algebra containing a two-dimensional superconformal algebra. For this reason, in order to fully specify the theory, including the insertion of extended operators, we have to supply additional data on top of the usual conformal data $\{\Delta_i , \lambda_{ijk}\}$ given by the scaling dimensional and OPE coefficients. More precisely, in the presence of a surface operator $V_\Sigma$ the remaining conformal symmetry constrains the one-point functions of local operators $\cO$ to the form,\footnote{When considering surface operators in flat space, the one-point function takes the more familiar form $\vev{\cO}_V~=~r ^{-2\Delta_\cO}b_\cO$, where $r$ is the transverse distance from the defect.}
\begin{equation}
\label{eq:local1pt}
    \vev{\cO}_V = b_\cO\,,    
\end{equation}
where we always take the operator to lie at the boundary of $\AdS_3\times S^1$ and normalise the one-point function so that $\vev{1}_V = 1$. 

In this work we will be particularly interested in the one-point function of the stress-tensor $T_{\mu\nu}$. As shown in \cite{Gomis:2007fi,Choi:2024ktc} this one-point function takes the form
\begin{equation}
\label{eq:T-one-pt}
    \vev{T_{ab}}_V = h \, g_{ab} \,,\qquad \quad \vev{T_{\psi\psi}}_V = -3h \,,
\end{equation}
where $g_{ab}$ is the metric on $\AdS_3$. The constant $h$ is sometimes dubbed the scaling weight of the surface operator, in analogy with the scaling dimension for local operators. The gauge theory defined on $\AdS_3\times S^1$ has a $\UU(1)$ spacetime symmetry that acts by rotating the $S^{1}$. Its associated conserved current $J_{\psi}$ acquires a non-zero one-point function which by supersymmetry can be related to $h$ as 
\begin{equation}
    \vev{J_{\psi}}_\Sigma = -3h \, \dd \psi\,.
\end{equation}
It is evident that the stress-energy one-point function, as defined in \eqref{eq:T-one-pt} is traceless, $\vev{T^{\mu}{}_{\mu}}_{\Sigma} = 0$, as required by classical conformal invariance. However, since we are considering an even-dimensional field theory with the insertion of an even-dimensional surface operator, typically there are conformal anomalies spoiling the tracelessness quantum mechanically. On top of the usual four-dimensional conformal anomalies, these include defect conformal anomalies along $\Sigma$. In the $2d-4d$ setup under consideration, such defect anomalies were analysed in \cite{Berenstein:1998ij,Graham:1999pm,Henningson:1999xi,Gustavsson:2003hn,Asnin:2008ak,Schwimmer:2008yh,Cvitan:2015ana,Jensen:2013lxa,Jensen:2019ano,Chalabi:2020iie,Chalabi:2021jud,Chalabi:2022wey} where it was shown that the parity even Weyl anomaly depends on three defect central charges and is given by
\begin{equation}\label{anomaly_exp}
    \vev{T^{\mu}{}_{\mu}} \supset -\delta(\Sigma)\frac{1}{24\pi} (b \bar{R} + d_{1} {\Pi}^{2} - d_{2}W_{ab}{}^{ab})\,.
\end{equation}
In this expression, $\bar{R}$ is the intrinsic scalar curvature of $\Sigma$, ${\Pi}^{m}_{ab}$ is the traceless second fundamental form and $W$ is the pull-back of the bulk Weyl tensor to $\Sigma$. Computing these quantities, one can check that the anomaly indeed vanishes for the defects considered above. Since we are studying conformal defects, one can relate the $d_2$ anomaly to the scaling weight of the conformal surface operator \cite{Chalabi:2022wey},
\begin{equation}
\label{eq:d2_centralcharge}
    h = -\frac{1}{36 \pi^{2}}d_{2}
\end{equation} 
Moreover, assuming that the average null energy condition (ANEC) holds in the presence of a defect, the above identity implies the bound $d_{2} \leq 0$ \cite{Jensen:2019ano,Capuozzo:2023fll}. In addition, the defect central charge $d_1$ can be related to the displacement operator two-point function \cite{Chalabi:2022wey}, which in turn is related to the scaling weight of the operator \cite{Bianchi:2019sxz}. Combining these results one finds (for defects preserving at least $\cN=(0,2)$ superconformal symmetry)
\begin{equation}
     d_1 = d_2 \,.
\end{equation}
Finally, we can deduce the $b$ central charge from the defect entanglement entropy \cite{Casini:2011kv,Jensen:2013lxa,Jensen:2019ano}. In particular, computing the entanglement entropy of a sphere of radius $l$ centred on the defect, one can define the defect entanglement entropy as
\begin{equation}
    S_{\rm EE}^{\rm def} = S_{\rm EE}^{\Sigma} - S_{\rm EE}^{\rm vac}
\end{equation}
where $S_{\rm EE}^{\rm vac}$ and $S_{\rm EE}^{\Sigma}$ are respectively the entanglement entropy computed in the vacuum state or the vacuum with the insertion of the surface operator. Following \cite{Jensen:2019ano}, the defect entanglement entropy of a surface operator in a four-dimensional bulk theory can be related to a linear combination of the defect central charges,
\begin{equation}\label{eq:EE_central}
    S_{EE}^{\mathrm{def}} \Big|_{\mathrm{log}} = \frac{1}{3} \left[ b + \frac{1}{3} d_{2} \right] \ln{\left(\frac{R}{\epsilon}\right)}
\end{equation}
where $\epsilon$ is a UV cut-off. 

The scaling weight and entanglement entropy the $\f12$-BPS Gukov-Witten surface operators introduced above have been computed respectively in \cite{Drukker:2008wr} and \cite{Gentle:2015ruo}, giving
\begin{align}
\label{eq:scaling_fieldtheory}
    h =&\, - \frac{1}{12\pi^{2}} \left(N^2-\sum_k  N_k^2\right) - \frac{2N}{3\lambda} \sum_{k=1}^{M}  N_{k}  (\beta_{k}^{2} + \gamma_{k}^{2})  \,,\\
    \label{eq:b_fieldtheory}
    b =&\, 3\left( N^2 - \sum_k N_k^2 \right)\,, 
\end{align}
Where $\lambda = g_{YM}^{2} N$. Note that $b$ in this expression is independent of the marginal parameters $\lambda$ and $\beta_i$, $\gamma_i$, as expected for an $A$-type anomaly.

The more general $\f14-$ and $\f18-$BPS surface operators are much less studied, and to the authors' knowledge the analogous defect observables are not available in the literature. In the following sections, we will recover the $\f12$-BPS observables, but more importantly, our holographic analysis allows us to go further and provide concrete predictions for a class of $\f14$- and $\f18$-BPS surface operator observables.

\section{Holographic defect solutions}
\label{sec:solutions}

In the remainder of this work we will propose a holographic description of the (generalised) Gukov-Witten defects introduced above. Since these defects in general have large scaling weight, the probe approximation is not valid. In order to overcome this we will construct novel backreacted solutions corresponding to these defects. We proceed in two steps, first describing the solutions in maximal gauged supergravity in five dimensions and in the next section uplifting and analysing them in type IIB supergravity. 

\subsection{The gauged supergravity theory}

Since we are studying solutions dual to surface operators in $\cN=4$ SYM the natural place to look for them is the maximal $\SO(6)$ gauged supergravity in five dimensions \cite{Gunaydin:1984qu,Pernici:1985ju,Gunaydin:1985cu}. Luckily, the backgrounds of interest in this work can all be constructed in a truncated $\cN=2$ supergravity theory which retains the $\cN=2$ graviphoton, two additional vector multiplets and four hypermultiplets. The truncation used in this work was derived in \cite{Liu:2007rv} and restricts to metric and axio-dilaton fluctuations in type IIB supergravity. In fact, all our backgrounds will have constant axio-dilaton hence we will effectively only consider three hypermultiplets. On top of the metric, this truncation contains three abelian vectors $A^{(I)}$, two neutral scalars, $\varphi_i$ and three charged scalars $\zeta_I= Y_I \e^{2\pi \ii \vartheta_I}$. It will be useful to trade the two unconstrained scalars $\varphi_i$ for three constrained scalars $X_I$ satisfying $\prod_I X_I = 1$,
\begin{equation}
    X_1 = \e^{-\f{\varphi_1}{\sqrt 6}-\f{\varphi_2}{\sqrt 2}}\,,\qquad X_2 = \e^{-\f{\varphi_1}{\sqrt 6}+\f{\varphi_2}{\sqrt 2}}\,,\qquad X_3 = \e^{\f{2\varphi_1}{\sqrt 6}}\,.
\end{equation}
When $Y_I=0$, this model reduces to the STU model (see e.g. \cite{Behrndt:1998ns}) while the extra scalars $Y_I$ reside in the three additional hypermultiplets. This truncation should be contrasted with the truncation \cite{Khavaev:2000gb,Bobev:2010de} used in \cite{Arav:2024exg} which contains the same amount of scalars but parametrises fluctuations of the IIB fluxes. These two truncations have a distinct physical interpretation. While our scalars $Y_I$ are dual to bosonic bilinears, the hypermultiplet scalars in \cite{Khavaev:2000gb,Bobev:2010de} correspond to fermionic bilinears. The gauged supergravity Lagrangian can be written as
\begin{equation}
    \label{eq:sugra-Lag}
    \begin{aligned}
        \cL =&\, \sqrt{|g|}\Big[R - \f12 X_I^{-2}|\dd X_I|^2 -\f12 |\dd Y_I|^2- \f14  X_I^{-2} |F^{(I)}|^2- 2m^2 \sinh^2 Y_I |D\vartheta_I|^2-V\Big]\\
        &\quad+ F^{(1)}\wedge F^{(2)}\wedge A^{(3)}\,, 
    \end{aligned}
\end{equation}
where $D\vartheta_I = \dd\vartheta_I+A^{(I)}$. The potential $V$ can be written in terms of a real superpotential $W$ as follows,
\begin{equation}
    \begin{aligned}
        W=&\, m \sum_I X_I\cosh Y_I\,,\\
        V =&\, 2\left[ \left(\partial_{Y_I} W\right)^2 + \right(\partial_{\varphi_1} W\left)^2+\right(\partial_{\varphi_2} W\left)^2-\f23 W^2 \right]\,,\\
        =&\, 2m^2 \left[ \sum_I X_I^2 \sinh^2 Y_I - 2 \sum_{I<J} X_IX_J \cosh Y_I \cosh Y_J \right]\,.
    \end{aligned}
\end{equation}
In the remainder of this work we fix $m=\f 2{L_{\AdS}}$ where $L_\AdS$ is the $\AdS$ length scale of the $\AdS_5$ vacuum solution.

\subsection{The backgrounds}

The surface operators of interest are (super)conformal, i.e. the states with the insertion of a surface operator preserve a two-dimensional superconformal algebra along the world-volume of the defect. Moreover, most of the defects of interest in this work will preserve rotations transverse to the defects as well. This leads us to the following ansatz to construct the relevant backgrounds,
\begin{equation}
\label{eq:ansatz}
    \begin{aligned}
        \ds_5^2  =&\, g_0(w) \ds^2_{\AdS_3} + g_1(w) \dd w^2 + g_2(w)\dd z^2\,,\quad & A^{(I)} =&\,  a_I(w) \dd z\,,\\
        X_I = &\, X_I(w)\,,&
        \zeta_I =&\, Y_I(w)\e^{2\pi\ii \,\vartheta_{I}}\,. 
    \end{aligned}
\end{equation}
where the coordinate $z\sim z+ 2\pi \Delta z$ parametrises a circle. Since the $\vartheta_I$ are linear in $z$ and always appear in the combination $D_\mu\vartheta_I$ we can always remove the non-integer part of $\vartheta_{I}$ by a gauge transformation of $A^{(I)}$.\footnote{In the presence of conical defects or multiple coverings one has to be more careful as some fractional shifts can only be removed by large gauge transformations and might therefore contain physical information \cite{Ferrero:2021etw,Bomans:2024mrf}.} Note however that this fixes the gauge and makes the value of $A^{(I)}$ at the boundary a physical observable.

Solutions satisfying the requirements above are in fact already available in the literature, and can be obtained as a double analytic continuation of the bubbling solutions of \cite{Chong:2004ce,Liu:2007rv}. These backgrounds are given as follow,
\begin{equation}
\label{eq:spindlesol}
    \begin{aligned}
        \ds_5^2  =&\, H(w)^{1/3} \left( \ds^2_{\AdS_3} + \ds^2_\Sigma \right)\,, \qquad & \ds_\Sigma^2 =& \,\f1{4f(w)}\dd w^2 + \f{f(w)}{H(w)}\dd z^2\,,\\
        A^{(I)} =&\, \left(\f{w}{h_I(w)}-1+\alpha_I\right)\dd z\,,&  X_I = & \,\f{H(w)^{1/3}}{h_I(w)}\,,\\
        \cosh Y_I =&\, h_I^\prime(w)\,. & \vartheta_{I} =&\, (1-\alpha_I)z\,, 
    \end{aligned}
\end{equation}
where the functions $f$ and $H$ are defined as
\begin{equation}
    f = H(w)-w^2\,,\qquad \qquad H(w)=\prod_I h_I(w)\,.
\end{equation}
The domain of the coordinate $w$ is non-compact and takes values in $w\in (w_0,\infty)$. These backgrounds are entirely determined by the three functions $h_I$, satisfying the following non-linear system of ODEs,
\begin{equation}
    \label{eq:ODEs}
    h_I^{\prime\prime}(w) = \f{H(w)}{h_I f(w)}\left( 1- (h_I^\prime(w)^2) \right)\,.
\end{equation}
Unfortunately, the complexity of solving this non-linear system of ODEs prohibits us from finding a general solution. For generic points in the parameter space we are therefore restricted to using numerical methods similar to the approach used in \cite{Liu:2007xj,Bomans:2023ouw}. Limiting the parameter space, a variety of analytic solutions can be found. Setting $Y_I=0$, the right hand side of the ODEs vanishes, allowing us to find the following simple solutions 
\begin{equation}
\label{eq:K0sol}
    h_I(w) = q_I + w\,.
\end{equation}
In this case the solution reduces to the local spindle background described in \cite{Ferrero:2020laf}. In contrast to \cite{Ferrero:2020laf}, we are interested in holographic defect solutions. For this reason we look at a different global completion of the local solution \eqref{eq:spindlesol}, where the coordinate $w$ takes values in the non-compact interval $w\in (w_0,\infty)$, with $w_0$ the largest real root of the function $f$.

\begin{figure}[!htb]
    \centering
    \begin{tikzpicture}
        \node (myPic) at (0,0) {\includegraphics[width=0.65\textwidth]{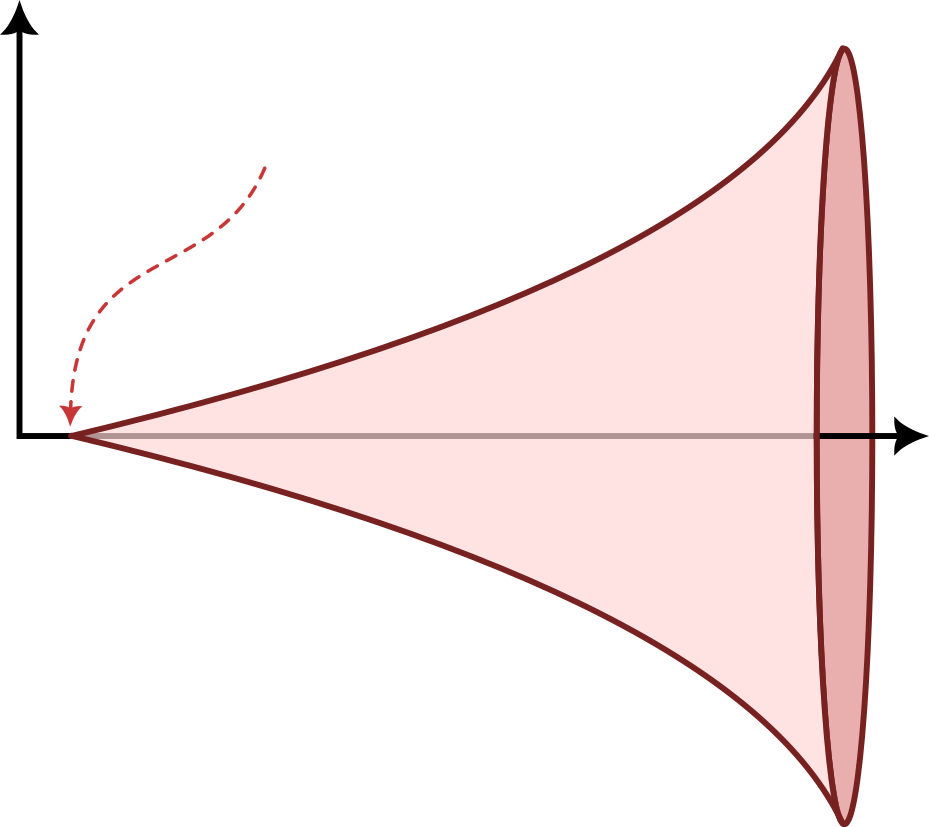}
	};
        \node at (5.2,3.8) {\Large$\AdS_5$};
        \node at (-1.8,3.2) {\Large$\AdS_3\times X$};
        \node at (-5.8,4) {\Large$f(w)$};
        \node at (-4.3,-0.7) {\Large$w_0$};
        \node at (5.2,0.2) {\Large$w$};
    \end{tikzpicture}
    \caption{The holographic defect solutions interpolate between asymptotically $\AdS_5$ and $\AdS_3\times X$. The precise form of the singularity $X$ depends on the type of defect.}
    \label{fig:enter-label}
\end{figure}
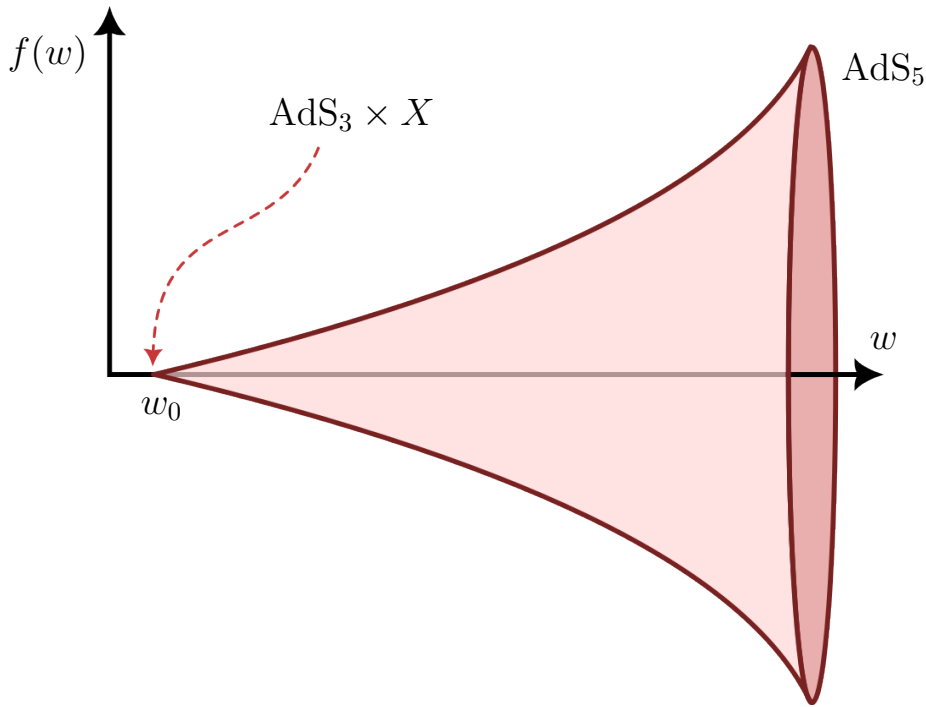

Note that another explicit solution can be obtained by setting $h_2=h_3=w$. In this case we can analytically solve the ODE for $h_1$ with solution%
\footnote{Clearly, there is a second solution to the ODE \eqref{eq:ODEs} with a minus sign in front of the square root. However, this solution does not asymptote to $\AdS_5$ hence we do not consider it further.}
\begin{equation}
\label{eq:Knot0sol}
    h_1(w) = 1 +\sqrt{(w+q_1-1)^2-K^2}\,. 
\end{equation}
which in the limit $K\to 0$ reduces to the spindle solution. However, note that this is a singular limit, as the derivative of $f$ at the simple root diverges. Around the double root at $w=0$ on the other hand the solution behaves exactly as it did for the original solution \eqref{eq:K0sol}.

\subsection{Regularity}
\label{subsec:regularity}

Before we continue, let us discuss the possible singularities arising in these solutions in some more detail. In the large $w$ regime, all solutions introduced above asymptote to the $\AdS_5$ background dual to the vacuum state of $\cN=4$ SYM. In particular, in the coordinates used above, the boundary of $\AdS_5$ is given by $\AdS_3\times S^1$. Changing coordinates to the usual Fefferman-Graham radial coordinate $w = 1/r^2$, the asymptotic metric becomes
\begin{equation}
    \ds_5^2 = \f{1}{r^2}\left( \dd r^2 + \ds_{\AdS_3}^2 + \dd z^2 + \cO(r^2) \right)\,.
\end{equation}
The boundary is conformal to $\bR^{1,3}$ but it will be useful to describe the defect with this metric. Note, however, that to guarantee the absence of conical defects on $\bR^{1,3}$, we have to fix the periodicity of the $S^1$ to be $2\pi$, i.e. $\Delta z = 1$. Asymptotically, the gauge fields are trivial up to possible monodromies around the non-contractible $S^1$,\footnote{Note that the periodicity of the holonomies is not manifest in our solutions. This was discussed in more detail in Appendix F of \cite{Arav:2024exg} where the apparent lack of periodicity was shown to originate from the boundary conditions imposed by supersymmetry. In particular, the periodicity can be restored by carefully taking the charge of the Killing spinor under the R-symmetry into account.}
\begin{equation}
    {\rm hol}_\infty(A^{(I)}) \equiv \f{1}{2\pi}\oint_{w=\infty}A^{(I)} = \alpha_I\,.
\end{equation}
On the other end, as $w\to w_0$, the metric is degenerate since $f(w)\rightarrow 0$. Depending on the properties of $f$ various types of singularities are possible. Here we discuss their most important characteristics, deferring the full analysis to Appendix \ref{app:zeroes}. The precise form of the singularity at $w=w_0$ will depend on the type of root of $f$, which can be simple, second or third order.

\textbf{Simple root:} For generic values of the parameters $q_I$ in \eqref{eq:K0sol}, the largest root will be a simple root. In this case, we can approximate the function $f(w-w_0) \simeq (w-w_0) f^\prime(w_0)$. Inserting this expression and redefining the coordinate $\rho^2=\f{w-w_0}{f^\prime(w_0)}$, we can approximate the metric near the root as
\begin{equation}
    \ds_\Sigma^2 \simeq \dd\rho^2 + \f{f^\prime(w_0)^2}{w_0^2} \rho^{2} \dd z^2\,.
\end{equation}
Therefore we find that near the largest root the two dimensional metric represents a $\bR^2/\bZ_k$ orbifold, with 
\begin{equation}
\label{eq:orb-order}
    k = \f{|w_0|}{f^\prime(w_0)}\,.
\end{equation}
Our solutions therefore correspond to orbibundles over an (infinite) disc with a conical singularity at the origin. To fully specify the $\UU(1)^3$ orbibundle we have to define the action of the orbifold $(\bR^2 \times (\bC^*)^3)/\bZ_k$,
\begin{equation}
    \bZ_k : (z,x_1,x_2,x_3) \rightarrow \left( \e^{\f{2\pi\ii}{k}}z\,, \e^{\f{2\pi\ii m_1}{k}}x_1\,, \e^{\f{2\pi\ii m_2}{k}}x_2\,, \e^{\f{2\pi\ii m_3}{k}}x_3 \right)\,,
\end{equation}
where $m_i\in \bZ_k$. With this we have fully specified the bundle data. Note that in order to preserve $\cN=(2,0)$ supersymmetry we have to demand that the orbifold action $\Gamma \in \SU(4)$. Hence, we have to demand that $1 + \sum_i m_i \in k \, \bZ$. With extended $\cN=(2,2)$ or $(4,4)$ supersymmetry then one or two of the $m_i$ vanish.

\textbf{Double root:} In this case we have two possibilities. Either we have a double root at $w=0$ and a simple root at $w<0$ or we have a double root at $w=w_0$ and a simple root at $w<w_0$. In the former case we necessarily have to put $h_2=h_3=w$ and $h(0)>1$. In this case the metric near the double root at $w=0$ takes the form
\begin{equation}
\label{eq:cylinder-Case}
    \ds^2_\Sigma \simeq \dd x^2 + \left(h(0)-1\right)\dd z^2\,,
\end{equation}
where we introduced the coordinate $x$ through $w= \exp \left[2x\sqrt{h(0)-1}\right]$. This is the metric on a cylinder. However, notice that the warp factor $H(w)^{1/3}$ also shrinks near this point so that we can still think of the space as capping off at $w=0$. Note that in this case we have an additional holonomy,
\begin{equation}
    {\rm hol}_0(A^{(1)}) \equiv \f{1}{2\pi}\oint_{w=0}A^{(1)} = \alpha_I-1\,.
\end{equation}
When allowing for two or more non-trivial functions the double root can also occur at a non-zero value $w=w_0$. In all such cases the metric near the root reduces to the metric on the hyperbolic plane (compactified on a circle). This behaviour was already noticed as the limit of spindle solutions in \cite{Couzens:2022lvg,Bomans:2023ouw} but its interpretation remains unclear hence we refrain from discussing it further. We refer to Appendix \ref{app:zeroes} for more details. 

\textbf{Triple root:} Similar to the double root there are two options. When $h_1 = w + 1$, $h_2 = h_3 = w$ we find a the triple root is located at $w= 0$. Alternatively, with three non-vanishing parameters we also have the option to have a triple zero at a non-zero value $w=w_0$. In both these cases it is unclear what the physical interpretation of these singularities is hence we won't discuss them further in the main text and refer to Appendix \ref{app:zeroes} for more details.

\textbf{Comments on $Y^{(I)}$:} The above analysis is independent of whether the scalars $Y_I$ are vanishing or not. However, when two or more $h_I$ are non-trivial we do not have analytic solutions. Moreover, the analytic solution \eqref{eq:Knot0sol} does not have the appropriate behaviour near the root. For this reason we are limited to using numerical methods to analyse these cases. Turning on the additional scalars breaks some of the global symmetries including the rotation transverse to the defect, possibly to a discrete subgroup. It would be interesting to get better analytic control over these cases.

\subsection{Supersymmetry}

The solutions considered in this work preserve varying degrees of supersymmetry, depending on how many gauge fields are turned on \cite{Benini:2013cda,Couzens:2021tnv,Suh:2021ifj}. The presence or absence of $Y_I$s does not significantly change this discussion. Each $Y_I$ is associated to the breaking of a particular $\UU(1)$, so the only constraint we impose is that the $Y_I$s associated to R-symmetries vanish. To simplify the discussion we therefore ignore the $Y$s in the following discussion. Similarly, in general we are allowed to turn on flat connections for non R-symmetries without altering the discussion below.

For ease of exposition, we simply summarise the various cases here, while an in depth discussion of the BPS equations is deferred to Appendix \ref{app:EOM-BPS}. With vanishing $Y_I$, the preserved supersymmetry is entirely determined by the values of the parameters $q_I$. In the generic case, all $q_I$ are distinct and non-vanishing. In this case the solution preserves $\cN=(0,2)$ superconformal symmetry (or $\cN=(2,0)$ depending on the choice of chirality of the preserved supercharge). Specialising to the case with one vanishing parameter, the solutions preserve $\cN=(2,2)$ superconformal symmetry while the case with two vanishing parameters preserves $\cN=(4,4)$ superconformal symmetry.

In addition to the supersymmetry, all solutions with vanishing $Y_I$ preserve at least $\UU(1)^3$ global symmetry. if two of the parameters are equal, this symmetry in enhanced to $\UU(1)^2\times \SU(2)$, while with three equal parameters, the global symmetry is enhanced to $\UU(1)\times \SU(3)$.

Finally, the flat connections associated to R-symmetries have to vanish. In the generic case this means that we need to impose $\sum_I\alpha_I = 0$. In the $\cN=(2,2)$ case the R-symmetry is given in \cite{Couzens:2021tnv}. Lastly, when we preserve $\cN=(4,4)$ superconformal symmetry, the Cartan generators of the two $\SU(2)$ R-symmetries are given by the sum and difference of the $\UU(1)$ generators corresponding to the vanishing $q_I$'s. Choosing $q_2=q_3=0$, we have to impose $\alpha_2=\alpha_3=0$. 

\section{Uplift to type IIB supergravity}
\label{sec:uplift}

To analyse the solutions introduced above, it proves useful to uplift them to ten-dimensional type IIB supergravity. Using the uplift formulae derived in \cite{Cvetic:1999xp,Cvetic:2000nc} all solutions above can be uplifted to solutions of type IIB supergravity. In this section we mostly restrict to solutions with $Y_I=0$ and comment on the case with non-vanishing $Y_I$. A more detailed description is relegated to Appendix \ref{app:uplift}. The resulting ten-dimensional solutions have a constant dilaton and the only non-trivial RR field strength is $F_5$.

\subsection{Generic case}

For the generic $\cN=(0,2)$ case the uplifted metric takes the form
\begin{equation}
    \ds_{10}^2 = \Delta^{1/2} H^{1/3} \left[ \ds_{\AdS_{3}}^2 + \ds_{Y_7}^2 \right]\,,
\end{equation}
where the seven-dimensional metric on $Y_7$ is
\begin{equation}
    \ds_{Y_7}^2 = \ds_\Sigma^2 + \f{1}{\Delta \, H^{2/3}}\sum_{I=1}^3 h_I (\dd\mu_I^2 + \mu_I^2 D \phi_I^2)\,,
\end{equation}
where $D\phi_I = \dd\phi_I + A^{(I)}$ and all the $\phi_I$ are $2\pi$-periodic. The coordinates $\mu_I$ parametrise a two-sphere and are constrained as $\sum_I\mu_I^2 = 1$. Finally, we defined
\begin{equation}
    \Delta = \sum_I X_I\mu_I^2\,.
\end{equation}
In the asymptotic region $w\rightarrow \infty$, the ten-dimensional metric takes the form
\begin{equation}
    \ds_{10}^{2} = \frac{1}{\rho^{2}} \left( \dd\rho^{2} + \ds^{2}_{\AdS_{3}} + \dd z^{2} + \cO(\rho^{2}) \right) + \left( \ds_{S^{5}}^{2} + 2 \sum_{I} \mu_{I}^{2} \alpha_{I} \dd\phi_{I} \dd z + \cO(\rho^{2}) \right)
\end{equation}
where we adopted standard Fefferman-Graham coordinates by performing the coordinate change $w = 1/\rho^{2}$. As expected, the asymptotic metric reduces to the standard $\AdS_5 \times S^5$ metric, twisted by the flat connections $\alpha_I \dd\phi_I$.

Moreover, note that we have fixed the periodicity of the Reeb vector of $Y_7$ by demanding that the boundary metric is conformal to $\bR^{1,3}$ without conical defects. Hence, unlike the spindle solutions we typically find a remaining singularity in the bulk. Such singularities were also observed on the topological disc solutions of \cite{Couzens:2021rlk}. Alternatively, we could choose the periodicity such that the bulk solution is entirely smooth but this will introduce a conical singularity at the boundary. This once again corroborates the general idea put forward in this work, that disorder surface operators can be equivalently be described as a surface operator in flat space or by putting the theory on flat space with a conical deficit.

\subsection{\texorpdfstring{$\f12$}{1/2}-BPS defects}

Having discussed some overarching global properties of the generic solutions let us come back to the $\f12$-BPS defects. These come in two flavours, either with a simple root at $w=w_0 >0$ or a double root at $w=0$. We focus mainly on the case with a single root but will comment on the double root case. 

Starting with vanishing $Y_I$ and plugging in the explicit $\f12$-BPS solution \eqref{eq:spindlesol}, the metric simplifies significantly and can be written as,
\begin{equation}
\begin{aligned}
\label{eq:12BPSmet-NoY}
    \ds_{10}^2 = \, \sqrt{w(w+q_1\mu^2)} \bigg[ & \ds_{\AdS_3}^2 + \f{\mu^2}{(w+q_1\mu^2)}ds_{S^3}^2 + \f{1}{4f(w)}\dd w^2 + \f{f(w)}{H(w)}\dd z^2 \\
    &\,+ \f{1}{w(1-\mu^2)}\dd\mu^2 + \f{(q_1+w)(1-\mu^2)}{w(w+q_1\mu^2)}\left( \dd\phi_{1} + A_1 \right)^2 \bigg]\,,
\end{aligned}
\end{equation}
where we introduced the following explicit choice of constrained coordinates,
\begin{equation}
    \mu_1 = \sqrt{1-\mu^2} \,,\qquad \mu_2 = \mu \sin\psi\,,\qquad \mu_3 = \mu \cos\psi\,.
\end{equation}
In this metric we see an emerging $S^3$ in the internal space, giving rise to the $\SU(2)\times \SU(2)$ R-symmetry preserved on the $\cN=(4,4)$ surface operator. The full internal space can be regarded as a $S^1_z \times S^1_{\phi_1} \times S^3$ fibration over the two-dimensional space spanned by $(w,\mu)$. At large $w$, the fields fall off as described above and we recover the $\AdS_5\times S^5$ background of type IIB supergravity. Close to the root, the behaviour depends on the multiplicity of the root. Moreover, the monodromy parameters in our gauge are fixed by supersymmetry to be $\alpha_2 = \alpha_3 = 0$.

\begin{figure}[htb!]
    \centering
    \begin{tikzpicture}
    \draw[->] (0,0) -- (6,0) node[anchor=north] {$w$};
    \draw[->] (0,0) -- (0,4) node[anchor=east] {$\mu$};
    \draw	(0,0.5) node[anchor=east] {0}
		(0,3.5) node[anchor=east] {1}
		(1,0) node[anchor=north] {$w_0$};
    \node[rectangle,draw,color={rgb:red,1;green,2;blue,1},fill={rgb:red,1;green,2;blue,1},minimum width=4.99cm,minimum height=3cm,anchor=south west,opacity=0.3] (r) at (1,0.5) {};
    \draw[very thick,dark-green] (6,3.5) -- (1,3.5) -- (1,0.5);
    \draw[very thick,dark-red] (1,0.5) -- (6,0.5);
    \filldraw[fill=dark-green,color=dark-green] (1,3.5) circle (3pt);
    \draw	(3.5,0.5) node[anchor=south] {$S^3\rightarrow 0$}
		(1,2) node[anchor=west] {$L=k$}
            (3.5,3.5) node[anchor=south] {$L=0$}
            (6,2) node[anchor=west] {$\AdS_5 \times S^5$}
            (1,3.5) node[anchor=south] {$\bR^4/\bZ_k$};
\end{tikzpicture}
    \caption{At a simple root, the non-compact space $\cM_7$ takes the form of an $S^1\times S^1 \times S^3$ fibration over a two-dimensional base space. The $S^3$ shrinks at the bottom edge, while the $S^1_{\phi_1}$ shrinks as we approach the two green edges. At the top left corner the internal space can be described by $\cM_7 \simeq S^3 \times \bR^4/\bZ_k$.}
    \label{fig:regular-defect}
\end{figure}
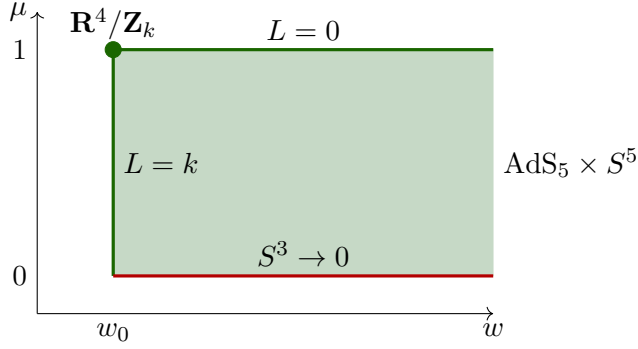

Considering the case with a simple root, it proves useful to rewrite the four-dimensional metric spanned by $(w,\mu,z,\phi_1)$ as a $S^1_z$ fibration over the space spanned by $(w,\mu,\phi_1)$. This simply amounts to completing the $\dd z^2$ in the metric above. The resulting metric can be written as
\begin{equation}
\begin{aligned}
\label{eq:12BPSmetric}
    \ds_{10}^2 = \sqrt{w(w+q_1\mu^2)}\bigg( & \ds_{\AdS_3}^2 + \f{\mu^2}{w+q_1\mu^2} \ds_{S^3}^2 \\
    &+ \f{\dd w^2}{4w^2(w-1+q_1)} + \f{\dd\mu^2}{w(1-\mu^2)}  + R_z (\dd z+ L \,\DD\phi_1)^2 + R_1 \DD\phi_1^2 \bigg)\,,
\end{aligned}
\end{equation}
where we defined
\begin{align}
    R_z =&\, \f{w+(q_1-1)\mu^2}{w+q_1\mu^2}
    \,,\nn\\
    L =&\, \f{1-\mu^2}{w+(q_1-1)\mu^2}\,, \\
    R_1 =&\, \f{(w-1+q_1)(1-\mu^2)}{w(w+1-q_1)\mu^2}\,.\nn
\end{align}
Observe that the radius $R_1$ vanishes both along $\mu=1$ and $w=w_0$, where one can easily check that it shrinks smoothly. In addition, $R_z$ vanishes only at the corner $(w,\mu) = (w_0,1)$. Finally, note that $S^3$ shrinks along the $\mu=0$ edge, once again in a completely smooth manner. Hence, the only possible singularity lies at the corner. This situation is sketched in Figure \ref{fig:regular-defect}. As indicated in this figure, the function $L$ is piecewise constant along the green edge, taking value $L=0$ at $\mu=1$ and value $L=k$ along the boundary with $w=w_0$, where $k=\f{1}{1-q_1}$ was defined in \eqref{eq:orb-order}. This jump indicates the presence of a monopole source for the $Dz$ fibration. To analyse the behaviour at the corner in more detail, we define the coordinates
\begin{equation}
    \mu = 1-\f{\sqrt{w_0}}{2}R^2 \cos^2\f\chi2\,,\qquad w = w_0+w_0^{3/2} R^2 \sin^2\f\chi2\,,
\end{equation}
and expand around $R\sim 0$. Near the corner we find that the metric reduces to
\begin{multline}\label{eq:R4Zk-uplift}
    \ds_{10}^2\simeq w_0^{1/2}\left( \ds_{\AdS_3}^2 + \ds^2_{S^3}\right) \\
    + \dd R^2 + R^2 \left( \f14 \left(\dd\chi^2 + \sin^2\chi \dd\phi_1^2\right) +\f{1}{k^2}\left( \dd z + \f k2 (1+\cos\chi)\dd\phi_1\right)^2\right)\,.
\end{multline}
In the second line we can recognise the line element of $S^3/\bZ_k$ written as a Hopf fibration. Combined with the radial $R$ direction, we obtain the metric on $\bR^4/\bZ_k$. Hence, in the case of a single root we find an orbifold singularity at the location of the monopole.

Next, let us consider the situation with the extra scalar $Y_1$ turned on. We do not have an explicit solution but we know that near the root we still have $h_1(w)\to 1$, while the ODE \eqref{eq:ODEs} shows that near the root $Y_1$ necessarily vanishes. For this reason, the behaviour near the root is largely unchanged. Similarly, near the asymptotic boundary the function $h_1$ necessarily becomes linear hence also there the leading asymptotic behaviour remains unchanged. Adding the scalar, the ten-dimensional metric becomes
\begin{equation}\label{eq:uplift_Ynot0}
\begin{aligned}
    \ds_{10}^2 = \sqrt{w(Z+\mu^2)}\bigg( & \ds_{\AdS_3}^2 + \f{\mu^2}{Z+\mu^2} \ds_{S^3}^2 \\
    &+ \f{\dd w^2}{4w^2(h_1(w)-1)} + \f{\dd\mu^2}{w(1-\mu^2)U} \\
    &+ R_z (\dd z+ L_{\phi_1} \,\dd\phi_1+ L_\mu \,\dd\mu)^2 + R_1 (\dd\phi_1+\wti L_\mu\dd\mu)^2\,, \bigg)\,,
\end{aligned}
\end{equation}
where we defined
\begin{equation}
\begin{aligned}
    R_z =&\, \f{Z}{Z+\mu^2}
    \,,&
    R_1 =&\, \f{(h_1(w)-1)(1-\mu^2)U}{w \,Z}
    \,.\\
    L_{\phi_1} =&\, \f{(1-\mu^2)U}{Z}\,, &
    L_{\mu} =&\, -\f{\sinh Y_1(w)\mu\sin 2\phi_1}{Z}\,, \\
    \wti L_{\mu} =&\, -\f{\sinh Y_1(w)\mu\sin 2\phi_1}{(1-\mu^2)U}\,,\qquad
\end{aligned}    
\end{equation}
in terms of the two functions 
\begin{align}
    U = \e^{Y_1(w)}\cos^2\phi_1 +\e^{-Y_1(w)}\sin^2\phi_1\,,\qquad
    Z = \mu^2(h_1(w)-1)+ w(1-\mu^2)U\,. 
\end{align}
The regularity analysis is very similar to the above. As $\mu\to 0$, the $S^3$ shrinks smoothly. Similarly, along the faces $\mu\to 1$ the $\phi_1$ cycle shrinks smoothly.\footnote{To see this is a bit subtle. One should make a change of coordinates $\mu= 1-r m(w,\phi_1)$ where $m$ can be defined implicitly as a PDE. Nonetheless, this suffices to show that the metric smoothly degenerates to $\bR^2$.} Next, let us consider the situation along $w=w_0$. At this point we find that the vector $V= \partial_z-w_0 \partial_{\phi_1}$ smoothly shrinks. Nonetheless, similar to the situation without $Y_1$, we find a remaining singularity at $w=w_0$ and $\mu=1$. where the metric degenerates to \eqref{eq:R4Zk-uplift}. This is not so surprising, since at this locus the additional scalar field vanishes.

Finally, let us briefly comment on the defect solution with a double root. This solution arises in the regime where $q_1 > 1$. Although the background reproduces various characteristics of a smeared brane distribution, after extensive analysis we could not massage it into a physically sensible brane-like singularity. We will leave an in-depth discussion of this case for future research.

\subsection{Comparison with bubbling surface operator backgrounds}

To better understand the backgrounds introduced above, let us compare them to known classes of defect solutions. In particular, we will show that the $\f12$-BPS backgrounds introduced above can be rewritten as bubbling surface operator backgrounds \cite{Drukker:2008wr}. The canonical form of a $\f12$-BPS solution in IIB was derived in \cite{Lin:2004nb}, where the form of the metric can be written as
\begin{multline}
    \label{eq:LLM-metric}
    ds^{2}_{\rm LLM} = y \sqrt{\frac{2a+1}{2a-1}} \dd s^{2}_{AdS_{3}} + y \sqrt{\frac{2a-1}{2a+1}} \dd s^{2}_{S^3} + \frac{2y}{\sqrt{4a^{2}-1}} (\dd \chi + V)^{2} \\
    + \frac{\sqrt{4a^{2}-1}}{2y} (\dd y^{2} + \dd x_{i}\dd x_{i})\,.
\end{multline}
Here $(x_{1},x_{2},y)$ are coordinates on the three-manifold $X$ over which the remaining directions are fibred, and $V$ is a one-form such that $dV = \f1y \star_3 \dd a$, with $\star_3$ denoting the Hodge dual on $X$. The function $a = a(x_{1},x_{2},y)$ fully determines the form of the metric -- as well as the five-form flux -- and is constrained to solve the PDE
\begin{equation}\label{eq:a_PDE}
    \partial_{i} \partial_{i} a + y \partial_{y} \left( \frac{\partial_{y} a}{y} \right) = \sum_{l=1}^{M} 2 \pi y_{l} \, \delta (y - y_{l}) \delta^{2} (\vec{x} - \vec{x_{l}})\,,
\end{equation}
The general solution to this equation is given by
\begin{equation}
    a(x_{1},x_{2},y) = \frac{1}{2} + \sum_{l=1}^{M} a_{l}(x_{1},x_{2},y)\,,
\end{equation}
with
\begin{equation}
    a_{l}(x_{1},x_{2},y) = \frac{(\vec{x} - \vec{x_{l}})^{2} + y^{2} + y_{l}^{2}}{2 \sqrt{((\vec{x} - \vec{x_{l}})^{2} + y^{2} + y_{l}^{2})^{2}-4y_{l}^{2}y^{2}}} - \frac{1}{2}\,.
\end{equation}
As shown in \cite{Gomis:2007fi,Drukker:2008jm} such backgrounds can be interpreted as holographic surface defects where the positions $(y_{l},\vec{x_{l}})$ appearing as sources on the right-hand side are related to the parameters $(\beta_{l},\gamma_{l})$ and to the choice of Levi subgroup $L = \prod_{l=1}^{M}U(N_{l})$. More precisely, we have
\begin{equation}
    y_{l}^{2} = \frac{N_{l}}{N}, \quad \vec{x_{l}} = 2\pi (\beta_{l},\gamma_{l})
\end{equation}
Moreover, the parameters $\alpha_{l}$, $\eta_{l}$ can be related to the periods of the 2-form potentials $B_{2}$ and $C_{2}$ of Type IIB supergravity,
\begin{equation}
    \alpha_l = - \f{1}{2\pi}\int_{\cC_l} B_2\,,\qquad \eta_l = - \f{1}{2\pi}\int_{\cC_l} C_2\,,
\end{equation}
where $\cC_l$ are non-trivial two-cycles constructed by fibring an $S^1$ over the interval connecting $y \in [y_l,\infty)$.

To compare our uplifted backgrounds to the discussion above, we can perform the explicit coordinate change
\begin{align}
    y =&\, \sqrt{w} \mu\,, \qquad\qquad \xi = \sqrt{(w + q_{1}-1)(1-\mu^2)}\,,\\
    \beta =&\, - \phi_{1}\,, \hspace{-0.8mm}\qquad\qquad \chi = z +\frac{1}{2}\frac{1}{1-q_{1}} \phi_{1}\,,\\
    a =&\, \frac{1}{2} + k \left[ \frac{\xi ^2+y^2+1-q_{1}}{2 \sqrt{\left(\xi ^2+y^2+1-q_{1}\right)^2-4 y^2 \left(1-q_{1}\right)}} - \frac{1}{2} \right]\,,\\
    V =&\, k \frac{\xi ^2+y^2+q_{1}-1}{2\sqrt{\left(\xi ^2+y^2+1-q_{1}\right)^2-4 y^2 \left(1-q_{1}\right)}} d\beta\,,
\end{align}
where we introduced polar coordinates on $\xi \e^{\ii\beta}$ on the $(x_1,x_2)$ plane. Note that the $\xi$ coordinate is well-defined only when $w > 1-q_{1}$, consistent with having a simple root in the five-dimensional defect solution. 

Using this explicit embedding we can identify the parameters of our supergravity solutions \eqref{eq:spindlesol} with the bubbling surface defect parameters. For the solutions without $Y_I$ scalars we have,
\begin{equation}\label{eq:defect_dictionary}
    M = k \,, \qquad y_{l}^{2} = \frac{N_{l}}{N} = (1-q_{1}) = \frac{1}{k}\,, \qquad \vec{x_{l}} = 2\pi (\beta_{l},\gamma_{l}) = 0\,.
\end{equation}
These identities provide an explicit dictionary between supergravity and defect parameters in the $\frac{1}{2}$-BPS case. All our solutions have vanishing three-form field strengths. However, turning on non-trivial monodromies for the five-dimensional gauge fields we find
\begin{equation}
    \alpha_l = \alpha_1\,.
\end{equation}
This concludes the identifications and therefore allows us to claim a clear interpretation of the defect solutions as holographic Gukov-Witten surface defects. With non-vanishing $Y_I$ the axial symmetry is broken, potentially allowing non-zero values of $\beta_l$ and $\gamma_l$. However, the map to LLM coordinates in this case is much more complicated and we leave a full analysis for future work. A preliminary analysis indicates that five-dimensional supergravity solutions are not fine-grained enough to give rise to fully localised (non-axially symmetric) sources in ten dimensions.

\subsection{\texorpdfstring{$\f14$}{1/4}- and \texorpdfstring{$\f18$}{1/8}-BPS defects}

So far we considered the $\f12$-BPS uplifted solutions, which fall in the general classification of \cite{Lin:2004nb}. These solutions are completely determined in terms of a single function $a(x_i,y)$ satisfying a cylindrical Laplace equation. This PDE is well understood and, provided with the appropriate boundary conditions, gives us a complete understanding of holographic regular $\f12$-BPS surface operators in $\cN=4$ SYM. Reducing the supersymmetry to $\f14$- or $\f18$-BPS surface defects, the solutions fall into the classification of \cite{Kim:2005ez}. Embedding our solutions into this framework is a delicate task and the benefits of doing so are not entirely clear in this case hence we will not pursue this. These solutions are defined in terms of a system of non-linear PDEs on a six-dimensional space, which is highly non-trivial to solve. Starting from this classification it is therefore entirely unclear what the appropriate boundary conditions are to capture the generalised Gukov-Witten defects discussed in this work. In contrast, our five-dimensional construction allows us to construct a class of holographic surface defects naturally generalising the $\f12$-BPS solutions.

To gain further insight in these backgrounds it proves instructive to investigate the local geometry close to the conical defect. Similar to the $\f12$-BPS defects, the two-dimensional conical defect combines with the internal dimensions to form a higher dimensional orbifold. However, in this case the orbifold takes the form $\bR^6/\bZ_k$. In the $\f14$-BPS case, this was already observed in \cite{Suh:2021ifj}.

In more detail, we want to zoom in around the point where $\mu_{1,2}\rightarrow 0$ and $w\rightarrow w_0$ simultaneously. In order to do so it is helpful to introduce the coordinates
\begin{equation}
    \mu = \f{R \cos\chi}{\sqrt{h_2(w_0)\sin^2\psi+h_3(w_0)\cos^2\psi}} \,,\qquad\quad w = w_0 + \f{F'(w_0) R \sin\chi}{h_2(w_0)h_3(w_0)}\,.
\end{equation}
In these coordinates the singular locus is located at $R=0$. Expanding near this point, we can write the ten-dimensional metric as
\begin{equation}
    \ds_{10}^2 = \sqrt{h_2(w_0)h_3(w_0)}\left( \ds_{\AdS_3}^2 + \f{h_1(w_0)}{h_2(w_0)h_3(w_0)}\DD\phi_1^2 + \f{1}{h_2(w_0)h_3(w_0)}\ds_{\cM_6}^2 \right)\,.
\end{equation}
where the six-dimensional space $\cM_6$ is given by
\begin{equation}
    \ds_{\cM_6}^2 = \dd\rho_1^2 + \rho_1^2 \left( \dd\phi_1 + \f{m_1}{k}\dd z \right)^2 + \dd\rho_2^2 + \rho_2^2 \left( \dd\phi_2 + \f{m_2}{k}\dd z \right)^2 + \dd\rho_3^2 +\rho_3^2 \f{\dd z^2}{k^2} \,.
\end{equation}
In this metric we defined $\rho_i$ as
\begin{equation}
\begin{aligned}
    \rho_1 =&\, \f{\sqrt{h_2(w_0)}R \cos\chi \sin\psi}{\sqrt{h_2(w_0)\sin^2\psi+h_3(w_0)\cos^2\psi}} \,,\qquad\qquad \rho_3 = R\sin\chi\,,\\
    \rho_2 =&\, \f{\sqrt{h_3(w_0)}R \cos\chi \cos\psi}{\sqrt{h_2(w_0)\sin^2\psi+h_3(w_0)\cos^2\psi}}\,.
\end{aligned}
\end{equation}
Hence in both cases we see that the geometry is locally described by a $\bR^6/\Gamma$ singularity with orbifold action 
\begin{equation}
    \left(z_1,z_2,z_3\right)\rightarrow \left(\e^{\f{2\pi\ii m_1}{k}} z_1\,, \e^{\f{2\pi\ii m_2}{k}} z_2\,,\e^{\f{2\pi\ii}{k}} z_3 \right)\,.
\end{equation}
When $\cN=(2,2)$ supersymmetry is preserved, the supersymmetry condition in the five-dimensional solution imposes that $m_1+m_2+1 \equiv 0\mod k$, hence  $\Gamma$ is a finite subgroup of $\SU(3)$ and the orbifold action is uniquely defined by one $\bZ_k$ valued parameter $m_1$. This situation is entirely analogous with \cite{Bomans:2024mrf}. On the other hand, when only $\cN=(2,0)$ supersymmetry is preserved, the supersymmetry constraint imposes that $m_1+m_2+m_3+1 \equiv 0\mod k$. In this case $\Gamma$ is a finite subgroup of $\UU(3)\subset \SU(4)$ and the orbifold has two free parameters.

\section{Holographic defect observables}                     %
\label{sec:observables}                                      %

In the previous section we identified our defect solutions as holographically describing a class of generalised Gukov-Witten surface defects. In this section we corroborate this conjecture by computing a variety of holographic observables in the presence of the defect. For the class of $\f12$-BPS defects we recover the expected field theory results, while for the $\f14$- and $\f18$-BPS defects our computations provide new predictions. 

\subsection{Symmetries and anomalies}

In the defect setting, the bulk anomaly polynomial should be amended with a defect anomaly polynomial, localised at the location of the defect.
\begin{equation}
    \cI^{\rm bulk-defect} = \cI_6^{\rm bulk} + \delta(X_4)\,\cI_4^{{\rm defect}}\,,
\end{equation}
where in our case $\cI_6^{\rm bulk}$ is the six-form anomaly polynomial of 4d $\cN=4$ SYM and $X_4\subset X_6$ is an auxiliary four-dimensional manifold on which the defect anomaly polynomial lives. The defects carry localised symmetries whose anomalies can be conveniently collected in a four-form defect anomaly polynomial,
\begin{equation}
    \cI^{\rm defect}_4 = \f12 k_{ab} c_1(F^a)\wedge c_1(F^b) - \f{k}{24}p_1(R)\,,
\end{equation}
where the label $a$ runs over all flavour symmetries of the defect. In analogy to standard anomaly polynomials for two-dimensional SCFTs \cite{Benini:2013cda}, we define $b_R = 3k_{RR}$, $b_L = 3k_{LL}$ and $k \propto b_R - b_L$. In addition, the defect R-symmetries should not have mixed anomalies, implying that $k_{RF} = k_{LF}= k_{LR} = 0$, where $F$ denotes any flavour symmetry.

The bulk + defect anomaly polynomial can be obtained systematically from IIB extending the anomaly inflow methods developed in \cite{Bah:2020jas} to the bulk-to-defect inflow. This analysis employs an auxiliary eleven-dimensional manifold $Y_{11}$, realised as a $Y_7$ fibration over a closed four-manifold $X_4$, $Y_7\hookrightarrow Y_{11}\rightarrow X_4$. The manifold $X_4$ can be understood as a fibration over the external $\AdS_3$ space and is the space on which the defect anomaly polynomial lives. Globally, the seven-dimensional space is an $S^5$ fibration over an infinite disc $\disc$,
\begin{equation}
    S^5 \hookrightarrow Y_7 \rightarrow \disc\,.
\end{equation}
As detailed below, we can then obtain the bulk + defect anomaly polynomial by integrating a particular eleven-form $\cI_{11}$ over the internal space.
\begin{equation}\label{eq:anom-integration}
    \cI^{\rm bulk-defect}  = \int \cI_{11}
\end{equation}
This eleven-form can be related to the topological couplings in ten-dimensional type IIB supergravity,
\begin{equation}
    \cI_{11} = \dd \cI_{10}\,,\qquad \quad S_{\rm top.}^{\rm IIB} = \int \dvol_{10}\cI_{10}\,.
\end{equation}
In the absence of a fluctuating axio-dilaton, the eleven form $\cI_{11}$ was argued to be \cite{Bah:2020jas},
\begin{equation}
    \cI_{11} = \f12 E_5 \wedge \dd E_5 + E_5 \wedge \cF_3 \wedge \cH_3\,.
\end{equation}
where $\cF_3$, $\cH_3$ and $E_5$ are globally well-defined forms on the auxiliary space $X_4\times Y_7$. Restricting to the ten-dimensional fibres, these forms reduce to the type IIB fluxes $F_3$ and $H_3$ while we have $F_5 = E_5 + \star E_5$ on the fibres.

Unlike in \cite{Bah:2020jas}, the internal manifolds $Y_7$, corresponding the defect geometries studied in this work, are non-compact. We can overcome this by cutting a disc at infinity, corresponding to the asymptotic $\AdS_5$ region and a small disc near the origin where the singularities are located. In the asymptotic region, the ten-dimensional geometry is simply $\AdS_5\times S^5$, possibly with some flat connections for the flavour symmetries turned on. Hence, in this region the integration in Equation \eqref{eq:anom-integration} should be performed over the external $S^5$, resulting in the bulk anomaly polynomial of $\cN=4$ SYM. 

Having taken into account the asymptotic region we are left with a cylinder and the local puncture geometry. From here on we can simply follow \cite{Bah:2020jas} and integrate the anomaly polynomial over the seven-manifold $Y_7$ corresponding to the cylinder and local puncture geometry. Since the Euler characteristic of the cylinder vanishes it does not contribute to the defect anomaly polynomial and the full defect anomaly polynomial can be obtained from the local defect geometry $X_7$ as described in \cite{Bah:2020jas} for the $\f12$-BPS punctures in $\cN=4$ SYM.

In this case the local puncture geometry $Y_7$ is an $S^3$ fibration over a four-manifold $Y_4$, which in turn can be described as a circle fibration over $\bR^3$,
\begin{equation}
    S^3 \hookrightarrow Y_7 \rightarrow Y_4\,,\qquad S^1 \hookrightarrow Y_4 \rightarrow \bR^3\,.
\end{equation}
at the points where the $S^1$ fibre shrinks the fibration has a monopole source. The solutions we found correspond to $M = k$ coincident monopole sources at $(y,x_i)=(\sqrt{w_0},0)$, which we can interpret as coordinates on this three-dimensional space. More generally, the bubbling surface defect solutions of \cite{Gomis:2007fi,Drukker:2008wr}, have $p$ monopole sources at positions $(y_a,x_{i,a})$, $a=1,\dots,p$, where flux quantisation imposes $\sum_a N y_a k_a = N$, with $N y_a$ and $k_a$ integers. At the locations of the monopole source, the 4d space is locally $\bR^4/\bZ_{k_a}$. Hence, we immediately find the defect anomaly polynomial corresponding to the $\f12$-BPS defects considered in this work as \cite{Bah:2020jas},
\begin{equation}
    \cI_4^{\rm defect} = -\chi\left(\SO(4)\right) \sum_{a=1}^p l_a\left(N^2 y_a^2 - N y_{a-1}\right)\,,\qquad l_a = \sum_{b=a}^p k_b\,.
\end{equation}
In this expression, the four-form characteristic class $\chi\left(\SO(4)\right)$ is defined as
\begin{equation}
    \chi\left(\SO(4)\right) = \f1{4\pi^2}\f18 f_2 \wedge f_2\,, 
\end{equation}
where $f_2$ is a background gauge field for the $\SO(4)\simeq \SU(2)\times \SU(2)$ is the R-symmetry of the defect. As mentioned at the beginning of this subsection we therefore find the defect anomalies 
\begin{align}
    b_R = b_L =&\, 3\sum_{a=1}^p l_a\left(y_a^2 - y_{a-1}\right)\\
    =&\, 3N^2\left(1-\f{1}{k}\right)\,.
\end{align}
where the second line applies to the rectangular punctures resulting from our lower-dimensional supergravity solutions. This matches the field theory expression for $\f12$-BPS punctures and will be furthermore confirmed by the holographic stress tensor one-point function computation below.

As noticed in Section \ref{sec:QFT-defects}, the defects carry localised flavour symmetries on top of the R-symmetry. In order to observe their existence in string theory it is useful to study the orbifold singularities in some more detail. All the orbifolds in the $\f12$-BPS case are of the form $\bC^2 / \Gamma$, where $\Gamma$ is a discrete subgroup of $\SU(2)$. It is well-known that these singularities can be resolved through crepant resolutions. The blow up introduces a collection of resolution two-cycles, $\bC\bP^1_\alpha$, where $\alpha= 1 \,,\dots\,, \rank{\fg_\Gamma}$ where $\fg_\Gamma$ is the ADE Lie algebra associated to $\Gamma$. In particular $\fg_{\bZ_k} = A_{k-1}$. The intersection pairing of these cycles reproduces the Cartan matrix $\cC_{\alpha\beta}$ of $\fg_\Gamma$. To each of these cycles we can associate the Poincar\'e dual harmonic 2-forms $\omega_{(2)\alpha}$, which satisfy 
\begin{equation}
    \int_{\bC^2 / \Gamma} \omega_{2\alpha}\wedge\omega_{2\beta} = - \cC_{\alpha\beta}\,, \qquad\quad \alpha\,,\beta = 1 \,,\dots \,, \rank(\fg_\Gamma)\,.
\end{equation}
We can now form a collection of five-cycle $\cC^\alpha_5 = \bC\bP^1_\alpha \times S^3$ on which we can wrap D5-branes, resulting in particle states in the external space corresponding to the defect world volume. These states can be interpreted as the W-bosons corresponding to a $\fg_\Gamma$ flavour symmetry of the defect. Alternatively, we can expand the the type IIB six-form potential $C_6$ on the harmonic forms as
\begin{equation}
    C_6 = c_1^\alpha \wedge \omega_{(2)\alpha}\wedge \dvol_{S^3} + \cdots\,.
\end{equation}
We can interpret the resulting vectors $c_1^\alpha$ as the external background vector fields corresponding to the symmetry. Combining the flavour symmetries from the various monopole sources, this gives rise to the flavour symmetry \eqref{eq:localised-flavour}. Note however that with the form of the eleven-form $\cI_{11}$ proposed in \cite{Bah:2020jas}, it is unclear how the anomalies of these flavour symmetries can be obtained through anomaly inflow. We leave this as a puzzle for future work.\footnote{We thank Federico Bonetti for insightful discussions on this point.} 

The whole analysis performed in this subsection straightforwardly generalises to $\f14$- and $\f18$-BPS defects. One can construct the equivariant five-form $E_5$ to find the anomalies for flavour symmetries from isometries of the internal manifold. We will not pursue this here but instead compute these anomalies by computing the stress tensor one-point function in the presence of the defect. In addition, the less supersymmetric defects allow for more general orbifolds of the form $\bC^3/\Gamma$ where $\Gamma$ is respectively a finite subgroup of $\SU(3)$ or $\UU(3)$. In the case of the $\f14$-BPS defects, the situation is entirely analogous to the $\f14$-BPS defects arising in the seven-dimensional spindle backgrounds studied in \cite{Bomans:2024mrf}. We will not discuss these cases in more detail but refer the reader to  \cite{Tian:2021cif} for a discussion of the resolutions of these orbifolds from which the flavour symmetries can be read off. Further lowering supersymmetry the zoo of possible orbifolds becomes even more unwieldy but apart from difficulties in resolving these singularities, the analysis is entirely analogous.

\subsection{Renormalised Action}

Having discussed the symmetries and anomalies, a second interesting holographic observable is given by the on-shell action of the defect background. In the dual field theory this quantity is related to the partition function in the presence of the dual defect.

Inserting the equations of motion, we can rewrite the on-shell action as
\begin{equation}\label{eq:on-shell}
    S_{\mathrm{on-shell}} = -\f{1}{16\pi G_N} \int_{\mathcal{M}} d^{5}x \sqrt{|g|} \sum_{I=3}^{3}\left[\frac{1}{6} X_{I}^{-2}|F^{(I)}|^{2} + \frac{2}{3} V  \right] + S_{\rm GH}\,.
\end{equation}
where 
\begin{equation}
     S_{GH} = \f{1}{8\pi G_N}\int_{\partial \mathcal{M}} d^{4}x \sqrt{|\gamma|} K 
\end{equation}
is the usual Gibbons-Hawking boundary term. As usual, the on-shell action diverges and needs to be renormalised. Following \cite{Emparan:1999pm,deHaro:2000vlm,Skenderis:2002wp} we can regularise the action by adding a superpotential and curvature counterterm to the action,
\begin{equation}
    S_{W} = \f{1}{8\pi G_N}\int_{\partial\cM} \dd^4 x \sqrt{|\gamma|} W\,,\qquad S_{\rm curv.} = \f{1}{32\pi G_N}\int_{\partial\cM} \dd^4 x \sqrt{|\gamma|}  R[\gamma] \,,
\end{equation}
where $R[\gamma]$ is the boundary Ricci tensor. Adding these counterterms to the diverging action, we obtain a finite result for the on-shell action. Inserting our solution we find,
\begin{equation}
    S_{\rm ren.} = \f{1}{16\pi G_N}\left(2w_0 -\f34\right) \vol_{\AdS_3}\vol_{S^1}
\end{equation}
where $\vol_{\AdS_3}$ is the regularised $\AdS$ volume and $\vol_{S^n}$ the volume of the $n$-sphere. Subtracting the contribution from the $\AdS$ vacuum solution we find the remaining defect contribution
\begin{equation}
    S_{\rm defect} = \f{1}{8\pi G_N}\left(w_0 -1\right) \vol_{\AdS_3}\vol_{S^1}\,.
\end{equation}
The partition function in the presence of a defect  scales as
\begin{equation}
    \vev{1}_\Sigma = R^{b/3}\,,
\end{equation}
where $R$ here is the $\AdS_3$ length scale \cite{Drukker:2008wr,Choi:2024ktc}. Comparing this with our on-shell action we have
\begin{equation}
    \vev{1}_\Sigma = \e^{-S_{\rm defect}} = R^{\f{\pi}{2G_N}(1-w_0)}\,,
\end{equation}
where we used the regularised volume $\vol_{\AdS_3} = -2\pi \log R$. As we will see below this matches the expected value of $b$.

More generally, we can expand the solution around the asymptotic boundary to explicitly observe the dependence on the sources and vevs. Using Fefferman-Graham coordinates, the metric can always be rewritten as
\begin{equation}
\begin{aligned}
    \ds^2 =& \frac{\dd\rho^{2}}{4 \rho^{2}} + \frac{1}{\rho} g_{a b} \dd x^{a} \dd x^{b} \\
    g_{a b} =& g_{(0),a b} + g_{(2),a b} \rho + g_{(4),a b} \rho^{2} + h_{(4),a b} \rho^{2} \log \rho + \dots\,.
\end{aligned}
\end{equation}
Where $g_{(0),a b}$ is the metric on $\partial \mathcal{M}$. Similarly, we can expand the scalar and gauge fields as
\begin{equation}\label{eq:field_expansions}
\begin{split}
   \frac{1}{2} \log{X_{I}} &= (\chi_{1}^{I} + \chi_{2}^{I}\rho + \dotsm)\rho + (\zeta_{1}^{I} + \zeta_{2}^{I} \rho + \dotsm) \rho \log{\rho}\\
   A^{(I)} &= (\alpha_{0}^{I} + \alpha_{1}^{I} \rho + \alpha_{2}^{I} \rho^{2} + \alpha_{3}^{I} \rho^{2} \log{\rho} + \dotsm)dz
\end{split}
\end{equation}
Where the expansion coefficients can be related to the vevs and sources of the dual fields. We can find the appropriate coordinate transformation from our solution perturbatively around $\rho = 0$. Up to linear order in $\rho$ it takes the form,
\begin{align}\label{eq:FG_5D}
    w =  \frac{1}{\rho} &+ \frac{1}{2} - \frac{1}{3} (q_{1}+q_{2}+q_{3}) \\
    &+ \frac{\rho}{144} \left[ 8 (q_{1}^2 +q_{2}^2+q_{3}^2-q_{1} q_{2}-q_{1} q_{3}-q_{2} q_{3})-24 (q_{1}+q_{2}+q_{3}) -9 \right]+ \mathcal{O}(\rho^{2})\,,\nn 
\end{align}
Comparing with our solution we immediately see that there are no sources turned on for the scalar fields, while the monodromy parameters $\alpha_I$ provide sources for the gauge fields.

Finally, in the above we restricted to the case with vanishing $Y_I$. However, note that the analysis remains identical with non-vanishing $Y_I$. In particular, the same counterterms should cancel all divergences. However, unlike the inflow analysis above, the approach here requires having an explicit solution. For the explicit solution \eqref{eq:Knot0sol} we evaluated the on-shell action and found that it remains unchanged.

\subsection{Stress-Energy Tensor 1-pt function}

With the on-shell action at hand we can compute a variety of one-point functions by differentiating with respect to the relevant source and subsequently setting the source to zero \cite{deHaro:2000vlm,Bianchi:2002ren}. Schematically we have
\begin{equation}
    \vev\cO = \f{1}{\sqrt{|g^{(0)}|}}\f{\delta S_{\rm ren}}{\delta J}\Big|_{J=0}\,.
\end{equation}
In particular, we are interested in computing the stress tensor one-point function in the presence of a defect,
\begin{equation}
    \langle T_{ij} \rangle = - \frac{2}{\sqrt{|g_{(0)}|}} \frac{\delta S_{\rm ren}}{\delta g_{(0)}^{ij}}
\end{equation}
Evaluating the resulting expressions on our background yields 
\begin{equation}
    \langle T_{i j} \rangle dx^{i} dx^{j} = -\frac{1}{16} \left[ 1 - \frac{8}{3}(q_{1} + q_{2} + q_{3}) \right] ( ds_{AdS_{3}}^{2}-3 dz^{2})
\end{equation}
which is traceless, as expected for a conformal defect. The first term is the vacuum contribution, corresponding to $q_{I} = 0$. Subtracting the vacuum contribution, we find the defect contribution to be
\begin{equation}
    \langle T_{i j} \rangle \big|_{\Sigma} dx^{i} dx^{j} = \frac{1}{6} (q_{1} + q_{2} + q_{3}) ( ds_{AdS_{3}}^{2}-3 dz^{2})
\end{equation}
As discussed in Section \ref{sec:QFT-defects} the stress tensor one-point function is related to the scaling weight of the defect, which is given by
\begin{equation}
    h = \frac{1}{6}(q_{1} + q_{2} + q_{3})
\end{equation}
The scaling weight in turn is related to the anomaly coefficient $d_{2}$ through \eqref{eq:d2_centralcharge}. Assuming the ANEC holds in the presence of defects yields the condition $d_{2} \leq 0$, meaning that for a physical defect the parameters $q_{I}$ must obey the sign constraint
\begin{equation}
    q_{1} + q_{2} + q_{3} \geq 0
\end{equation}
It can be shown that this constraint is indeed obeyed by all $\f12$-BPS and $\f14$-BPS spindle solutions, while for the $\f18$-BPS solution with three independent parameters $q_I$ it is an additional condition that must be imposed. 

\subsection{Defect Entanglement Entropy}
\label{subsec:EE_calculation}

As a final defect observable we consider the defect entanglement entropy. This section closely follows the formalism of \cite{Estes:2014de} adjusted to our context. In a holographic setting the EE is computed using the Ryu-Takayanagi (RT) formula \cite{Ryu:2006}:
\begin{equation}
    S_{EE} = \frac{\mathcal{A}_{min}}{4 G_{N}}
\end{equation}
Where $\mathcal{A}_{min}$ is the area of the extremal surface on a fixed time slice that minimizes the bulk area functional, anchored to the entangling region on the boundary, and $G_{N}$ is the gravitational constant in the bulk theory. By using the Ryu-Takayanagi formula we can reduce the problem of calculating the entanglement entropy in the defect boundary theory to a purely geometric problem. 

We take the entangling region to be a Euclidean 3-ball $\mathcal{B}$ of radius $R$ centred on the defect. The bulk geometry belongs to a class of ten-dimensional metrics which can be written in the form
\begin{equation}
    \ds^2 = s(w,x^{a})^{2} \ds_{\AdS_{d}}^{2} + \rho(w,x^{a})^{2} \dd w^{2} + G_{bc}(w,x^{a}) \dd x^{b} \dd x^{c}\,,
\end{equation}
Where the $x^{a}$ are coordinates on a compact manifold $\mathcal{M}_{9-d}$ with metric $G_{ab}$, and the $\AdS$ component is written in Poincar\'e coordinates
\begin{equation}
    ds^{2}_{\AdS_{d}} = \frac{1}{u^{2}} (\dd u^{2} - \dd t^{2} + \dd r^{2} + r^{2}\dd\Omega_{d-3})
\end{equation}
With the defect located at the boundary $u = 0$. Parametrising the hypersurface as $u = u(r,w,x^{a})$, the area functional is given by:
\begin{equation}
\begin{split}
    \mathcal{A} = \vol_{S^{d-3}} \int& \dd x^{a} \dd r \dd w \sqrt{\mathrm{det}G} \\& \times r^{d-3} \rho \left( \frac{s}{x}\right)^{d-2} \sqrt{1 + (\partial_{r}x)^{2} + \frac{s^{2}}{x^{2}} \left( \frac{(\partial_{w}x)^{2}}{\rho^{2}}+G^{ab} \partial_{a} x  \partial_{b} x \right)}
\end{split}
\end{equation}
Without loss of generality we can pick a time slice $t = 0$. It was shown in \cite{Jensen:2013rga} that the extremal surface obtained by variations of the above functional satisfies $w^{2} + r^{2} = R^{2}$, and furthermore that this is a global minimum of the area. The minimal area is then given by
\begin{equation}
    \cA_{\rm min} = \vol_{S^{d-3}} R \int \dd x^{a} \dd u \dd w \sqrt{\det G} \rho s^{d-2} \frac{(R^{2}-x^{2})^{(d-4)/2}}{x^{d-2}}
\end{equation}
Note that the integral neatly splits into an integral over $u$ and an integral over $w$ and $\mathcal{M}_{9-d}$.

The solution describing a surface operator has $d = 3$, and the metric functions were found through the uplift in the previous section (see Section \ref{sec:uplift}). Note that for ease of computation in this section we redefine $\mu = \cos{\theta}$. Then, substituting and integrating over the angular coordinates $(z,\phi_{I})$ yields
\begin{equation}
    \mathcal{A}_{min} = 32 \pi^{4} R \int_{\epsilon_{x}}^{R} dx \frac{1}{x\sqrt{R^{2}-x^{2}}} \mathcal{I} = 32 \pi^{4} \left( \ln{\frac{2R}{\epsilon_{x}}} + \dotsm \right) \mathcal{I}
\end{equation}
The cut-off $\epsilon_{x}$ must be introduced to regulate the divergences caused by correlations between UV modes near the boundary of the entangling region $\partial \mathcal{B}$. Postponing momentarily the question of choosing appropriate integration bounds, the remaining integral $\cI$ is given by
\begin{equation}
    \cI = \frac{1}{4}\int \dd\theta \dd\psi \int \dd w f_{\AdS_{3}} f_{w} f_{\phi_{1}}f_{\phi_{2}}f_{\phi_{3}} \sqrt{(f_{\theta\psi}^{4} - 4 f_{\psi}^{2}f_{\theta}^{2}) \left( \frac{f_{z\phi_{1}}^{4}}{f_{\phi_{1}}^{2}}+\frac{f_{z\phi_{2}}^{4}}{f_{\phi_{2}}^{2}}+\frac{f_{z\phi_{3}}^{4}}{f_{\phi_{3}}^{2}} - 4f_{z}^{2} \right)}
\end{equation}
This integral looks rather intimidating, but plugging in the metric functions and simplifying yields the miraculously simple result
\begin{equation}\label{eq:int_partial}
    \cI = \frac{1}{2} \int \dd\theta \dd\psi \int_{w_0}^{\infty} \dd w \cos{\theta}^{3} \cos{\psi} \sin{\theta} \sin{\psi}\,,
\end{equation}
The $w$ integral is clearly divergent, as it runs from the root $w_{0}$ up to infinity. We must then introduce a second regulator located at $\rho = \epsilon_{\rho} \ll 1$, with $\rho$ the radial Fefferman-Graham coordinate. Since we are working in ten dimensions, the cut-off for the $w$ integral is really a hypersurface $\Lambda$, which we can obtain using a ten-dimensional Fefferman-Graham transformation between $\rho$ and $w$. The full transformation is discussed in Appendix \ref{app:uplift}. The resulting cut-off can be written as,
\begin{equation}\label{eq:lambda_cutoff}
\begin{split}
    \Lambda (\epsilon_{\rho},\theta,\psi) = \frac{1}{\epsilon_{\rho}} + \frac{1}{8} [& 4+2\cos ^2(\theta ) (q_{1}+q_{2}\cos^{2}(\psi )+q_{3}\sin^{2}(\psi ))\\&-4 (q_{1}+q_{2}+q_{3})+2 \sin ^2(\theta ) (q_{2}+q_{3})] + \dots\,.
\end{split}
\end{equation}
Performing the integration in \eqref{eq:int_partial} then yields
\begin{equation}
    \cI = \frac{1}{16} \left[\frac{1}{\epsilon_{\rho}} + \frac{1}{2} - \frac{1}{3} (q_{1}+q_{2}+q_{3}) - w_{0}\right] + \mathcal{O}(\epsilon_{\rho})
\end{equation}
The divergence in $\epsilon_{\rho}$ is eliminated by vacuum subtraction. Taking the limit $\epsilon_{\rho}\rightarrow 0$ we finally obtain:
\begin{equation}
    \mathcal{A}_{\min} - \mathcal{A}_{min,0} = 2\pi^{4} \left[ (1-w_{0}) - \frac{1}{3} (q_{1} + q_{2} + q_{3}) \right] \left( \ln{\frac{2R}{\epsilon_{x}}} + \dotsm \right)
\end{equation}
To wrap up the calculation, we use the Ryu-Takayanagi formula and compare the result with the expression for the universal part of $S_{EE}$ given in \eqref{eq:EE_central}. We find
\begin{equation}\label{eq:EE_expression}
    S_{EE}^{\mathrm{def}} \Big|_{\mathrm{log}} \equiv \frac{1}{3} \left[ b + \frac{1}{3} d_{2} \right] = \frac{\pi^{4}}{2 G_{N}^{(10)}}\left[ (1-w_{0}) - \frac{1}{3} (q_{1} + q_{2} + q_{3}) \right] 
\end{equation}
Where $G_{N}^{(10)}$ is the ten-dimensional gravitational constant.

Reintroducing $Y_1 \neq 0$, the uplifted ten-dimensional metric is given in \eqref{eq:uplift_Ynot0}. It can be verified that the metric functions still simplify, but now the introduction of the new scalar breaks rotational invariance over the $\phi_1$ coordinate. Therefore, $\cI$ is modified to become
\begin{equation}
    \cI = \frac{1}{4\pi} \int \dd \theta \dd \psi \dd \phi_1 \int_{w_0}^{\Lambda (\epsilon_\rho,\theta,\phi_1)} \cos^3 \theta \cos \psi \sin \theta \sin \psi
\end{equation}
Hence, the only difference from the previous calculation is the form of $\Lambda$. The appropriate Fefferman-Graham transformation is discussed in Appendix \ref{app:uplift}. We find
\begin{equation}\label{eq:lambda_cutoffY}
    \Lambda(\epsilon_\rho,\theta,\phi_1) = \frac{1}{\epsilon_\rho} + \f12 [2 + q_1 \cos^2 \theta - 2 q_1 + K \sin^2 \theta \cos 2 \phi_1] + \dotsm
\end{equation}
Clearly, the integration over $\phi_1$ makes the last term drop out, so we recover the same result as for the un-deformed case.

Summarising the above, and converting to field theory quantities, we find the $b$ and $d_2$ anomaly coefficients as
\begin{equation}
    b = 3 N^{2} (1-w_{0}), \quad d_{2} = - 3 N^{2} (q_{1} + q_{2} + q_{3})
\end{equation}
where we used the holographic relations 
\begin{equation}
    G_{N}^{(10)} = \frac{1}{2} \pi^{6} l_{P}^{8}\,,\qquad L_{AdS}^{4} = \pi N l_{P}^{4}\,,\qquad G_{N}^{(10)} = \vol_{S^5} G_{N}^{(5)}\,.
\end{equation}
It is now straightforward to compare these results with the expressions given in $\eqref{eq:scaling_fieldtheory}$ for the $\f12$-BPS surface defects. Using the explicit identification between supergravity and defect parameters obtained in \eqref{eq:defect_dictionary}, we verify that our results match the $\f12-$BPS field theory calculation. In addition, our holographic analysis offers a predictions for the values of $b$ and $d_2$ for a large class $\f14$- and $\f18$-BPS surface defects. 


\bigskip
\bigskip
\bigskip

\leftline{\bf Acknowledgements}
\smallskip
\noindent It is with pleasure that we thank Chris Couzens, Federico Bonetti, Jaume Gomis and Brandon Robinson for useful and inspiring discussions. In particular, we thank Chris Couzens for comments on the draft. The contributions of PB were made possible through the support of grant No. 494786 from the Simons Foundation (Simons Collaboration on the Non-perturbative Bootstrap) and the ERC Consolidator Grant No. 864828, titled “Algebraic Foundations of Supersymmetric Quantum Field Theory'' (SCFTAlg). LT is supported by a STFC Doctoral Training Partnership grant.

\newpage                                                       
\appendix

\section{Gauged supergravity solutions}
\label{app:EOM-BPS}

In this appendix we discuss in more detail the $5d$ gauged supergravity solutions described in the main text. As described in Section \ref{sec:solutions}, the relevant supergravity model is obtained as a $\UU(1)^3$ truncation of the full $5d$ $\cN=8$ $\SO(6)$ gauged supergravity.

\subsection{Equations of motion and BPS equations}

The bosonic equations of motion can be written as follows. The scalar equations of motion read
\begin{equation}
\begin{aligned}
    0=&\,\nabla^2 \varphi_1 - \f{1}{2\sqrt 6}\left( X_1^{-2} |F^{(1)}|^2 + X_2^{-2} |F^{(2)}|^2 - 2 X_3^{-2} |F^{(3)}|^2\right) - \partial_{\varphi_1} V\,,\\
    0=&\,\nabla^2 \varphi_2 - \f{1}{2\sqrt 2}\left(X_1^{-2} |F^{(1)}|^2 - X_2^{-2} |F^{(2)}|^2\right) - \partial_{\varphi_2} V\,,\\
    0=&\,\nabla^2 Y_I -4 \left|\dd\vartheta_I+ m A^{(I)}\right|^2\sinh 2Y_I - 2\partial_{Y_I} V\,,
\end{aligned}
\end{equation}
where $V$ is the scalar potential. The gauge field equations of motion are,
\begin{equation}
    \nabla^\mu \left(X_I^{-2}F^{(I)}_{\mu\nu}\right) = 4m D_\nu\vartheta_{I} \sinh^2 Y_I\,. 
\end{equation}
where we define $D_\mu \vartheta_I= \partial_\mu \vartheta_I + m A^{(I)}_\mu$. Finally, the Einstein equation can be written as
\begin{equation}
    \begin{aligned}
        0=& \, R_{\mu\nu} - \f12 \partial_\mu\varphi_i\partial_\nu\varphi_i- \f12 \partial_\mu Y_i \partial_\nu Y_i - 2 \sinh^2 Y_I D_\mu\vartheta_ID_\nu\vartheta_I \\
        &\,-\f12 X_I^{-2}\left( F^{(I)}_{\mu\alpha}F^{(I)\alpha}_{\nu} - \f16 |F^{(I)}|^2 g_{\mu\nu} \right) - \f{1}{3}g_{\mu\nu} V\,. 
    \end{aligned}
\end{equation}
The constant $m$ sets the scale of the $\AdS_5$ vacuum solution of the 5d supergravity model. In our conventions we have $L_{\AdS_5} = \f 2m$. Most of the time we choose to put $m=1$. 

The supersymmetry transformations of this supergravity model are given by \cite{Liu:2007rv}
\begin{align}
    \delta\psi_{\mu i} = &\, \left[\nabla_\mu - \f{\ii}{2} \cA_\mu + \f{\ii}{24}\left( \gamma_\mu{}^{\nu\rho} - 4\delta_\mu^\nu\gamma^\rho \right) \cF_{\nu\rho} + \f{1}{6} W\gamma_\mu \right]\epsilon\,,
    \label{eq:susy-vars-1}\\
    \delta\chi_i = &\, \left[ \slashed{\partial}\varphi_i +\f{\ii}{2} \sum_I \partial_{\varphi_i}(X_I)^{-1} \slashed{F}^I - 2 \, \partial_{\varphi_i}W \right]\epsilon\,,
    \label{eq:susy-vars-2}\\
    \delta\lambda_{I} = &\, \left[ -\ii \slashed{\partial}Y_I - 2 \slashed{D}\vartheta_I \sinh Y_I - 2 \ii\, \partial_{Y_I}W  \right]\epsilon\,,\label{eq:susy-vars-3}
\end{align}
where we define
\begin{equation}
    \cA_\mu = \sum_I D_\mu\vartheta_I \cosh Y_I \,,\qquad \cF_{\mu\nu} = \sum_I X_I^{-1} F_{\mu\nu}^{(I)}\,.
\end{equation}
And the superpotential $W$ is
\begin{equation}
    W = m \sum_I X_{I} \cosh Y_I\,.
\end{equation}
%

\subsection{Supersymmetry of the solutions}

In the following we show that the generic solutions discussed in the main text preserve $\f18$ of the total $\cN = 8$ supersymmetries of $5d$ $SO(6)$ supergravity. For a more detailed discussion we refer to \cite{Suh:2021ifj} and to the appendices of \cite{Ferrero:2021etw}, which give an explicit derivation of the solution and of the spinors $\epsilon_i$ in the un-deformed case ($Y_I = 0$).

We begin by assuming the ansatz given in \eqref{eq:ansatz}, which has metric
\begin{equation}
    \ds_{5}^{2} = g_{0}(w) \ds_{AdS_{3}}^{2} + g_{1}(w) \dd w^2 + g_{2}(w) \dd z^2\,,
\end{equation}
while the gauge fields take the form
\begin{equation}
    A^{(I)} = A_{z}^{(I)}(w) \dd z\,.
\end{equation}
The scalar fields are assumed to be functions only of $w$. Note that in our truncation $X_{1}X_{2}X_{3} = 1$. 

We rewrite the spinor supersymmetry parameters $\epsilon$ as
\begin{equation}
    \epsilon = \theta \otimes \eta\,,
\end{equation}
where $\theta_i$ are two 2-component $AdS_3$ Killing spinor satisfying
\begin{equation}
    \nabla_a^{AdS_{3}} \theta = \f12 \mathbf{k} \, \beta_a \theta\,,
\end{equation}
with $\mathbf{k} = \pm 1$ for respectively a chiral or anti-chiral spinor, $a = 0,1,2$ and $\beta_a$ are $3d$ gamma matrices. We make the choice $\beta_0 = i \sigma_2$, $\beta_1 = \sigma_1$, $\beta_2 = \sigma_2$. The quantity $\eta$ is a spinor on the surface parametrized by $w$ and $z$,
\begin{equation}
    \eta = \begin{pmatrix} \eta_1 (w,z) \\ \eta_2 (w,z) \end{pmatrix}\,.
\end{equation}
We pick the following five-dimensional gamma matrices
\begin{equation}
    \gamma_a = \beta_a \otimes \sigma_3 \, , \quad \gamma_3 = \mathbf{1} \otimes \sigma_1\, , \quad \gamma_4 = \mathbf{1} \otimes \sigma_2\,,
\end{equation}
and use them to rewrite the BPS conditions as equations on the two-component spinor $\eta$. The gaugino equation \eqref{eq:susy-vars-2} becomes
\begin{equation}
    0= \left[ g_1^{-1/2} \varphi_i' \,\sigma_1 - (g_1 g_2)^{-1/2} \sum_I\partial_{\varphi_i}(X_I^{-1})F_{wz}^I \,\sigma_3   -2\, \partial_{\varphi_i}W\right] \eta\,,
\end{equation}
where here and in the following a prime denotes differentiation by $w$. The $\AdS$, $w$ and $z$ components of \eqref{eq:susy-vars-1} yield respectively
\begin{align}
    0 =&\, \frac{1}{4} \frac{g_0'}{g_0} g_1^{-1/2} \left(\sigma_1 \eta \right) + \f12 \left(g_0^{-1/2} - \f16 (g_1 g_2)^{-1/2} \sum_I \frac{A_{z}^{(I) \prime}}{X_I} \right) \left( \sigma_3 \eta \right) + \frac{1}{6} W \eta\,, \label{eq:BPSgrav1}\\
    0 =&\, \partial_w \eta - \frac{\ii}{6} g_2^{-1/2} \sum_I \frac{A_{z}^{(I) \prime}}{X_I} \left( \sigma_2 \eta \right)+ \f16 g_1^{1/2} W \left(\sigma_1 \eta\right)\, , \label{eq:BPSgrav2}\\
    0 =&\, \partial_z \eta -\frac{\ii}{4} g_2' g_1^{-1/2} g_2^{-1/2} \left(\sigma_3 \eta \right) +  \frac{\ii}{6} g_1^{-1/2} \sum_I \frac{A_{z}^{(I) \prime}}{X_I} \left( \sigma_1 \eta \right) \nonumber \\
    & \qquad\qquad - \frac{\ii}{2} \sum_I D_{z} \vartheta_I \cosh Y_I \eta+ \f16 g_2^{1/2}  W \left( \sigma_2  \eta \right)\,. \label{eq:BPSgrav3}
\end{align}
Finally, the variation \eqref{eq:susy-vars-3} reduces to the equation
\begin{equation}\label{eq:varYIequation}
    \left[g_1^{-1/2} Y_{I}^{\prime} \,\sigma_1 -2\ii\, g_2^{-1/2} D_z\vartheta_I \,\sigma_2   + 2 \,\partial_{Y_I}W \right] \eta = 0\,.
\end{equation}
Now we substitute in the explicit solution given in \eqref{eq:spindlesol}. It is easy to show that equations \eqref{eq:susy-vars-2} and \eqref{eq:susy-vars-3} can be rewritten as a single projection condition on $\eta$ of the from
\begin{equation}\label{eq:BPSprojector}
    \cP \eta = - \ii \eta\,,
\end{equation}
with the projector
\begin{equation}
    \cP = \ii H^{-1/2} \begin{pmatrix}
        w & f^{1/2}\\
        f^{1/2} & -w
    \end{pmatrix}\,,
\end{equation}
and $\cP^2 = -1$. This projector halves the number of independent real supercharges. Therefore, solutions of this form will preserve $\f12$ of the $\cN = 2$ supersymmetries, or $\f18$ of the full $\cN = 8$ supersymmetries. It remains to be checked what additional constrains on $\eta$ are imposed by the other BPS equations, and whether they are consistent with this interpretation. The $\AdS$ component \eqref{eq:BPSgrav1} may be rewritten as
\begin{equation}
    -\frac{1}{6} \ii m W \left( \cP  + \ii \right) \eta = 0\,.
\end{equation}
And is therefore immediately satisfied given \eqref{eq:BPSprojector}. Turning to the $z$ equation \eqref{eq:BPSgrav3}, we may rewrite it in the form
\begin{equation}
     \left[ \partial_z  + \xi \cP  + \f16 \left( 2 \sum_{I} A_z^{(I)} + 2 (\xi - 1) + \frac{H'}{\sqrt{H}} \sigma_3 \right)\left( \cP + \ii \right) \right] \eta = 0\,,
\end{equation}
where $\xi = (\alpha_1 + \alpha_2 + \alpha_3 - 1)/2$. Using \eqref{eq:BPSprojector} this reduces to
\begin{equation}
    \partial_z \eta = \ii \xi \eta\,.
\end{equation}
This equation can be solved by setting
\begin{equation}\label{eq:eta_hat}
    \eta (w,z) = e^{i \xi z} \hat{\eta}(w) = e^{i \xi z} \begin{pmatrix} \hat{\eta}_1 (w) \\ \hat{\eta}_2 (w) \end{pmatrix}\,.
\end{equation}
Then, substituting \eqref{eq:eta_hat} into equation \eqref{eq:BPSgrav2} and using \eqref{eq:BPSprojector} we obtain a linear ODE for $\hat{\eta}_1$, which can be solved to give
\begin{equation}
    \hat{\eta}_1 = \frac{\sqrt{H^{1/2}-w}}{H^{1/6}} \, , \quad \Rightarrow \quad \hat{\eta}_2 = -\frac{\sqrt{H^{1/2}+w}}{H^{1/6}} \,,
\end{equation}
up to constant prefactors. Finally, we turn to the supersymmetry variation of the new scalar $Y_I$ \eqref{eq:varYIequation}. It can be verified that the $Y_1$ equation may be written as
\begin{equation}
    \frac{2 K H^{1/3}}{f^2} \left( 1 - h_1^{-1/2} \sigma_3 \right) \left( \cP + \ii \right) \eta = 0\,.
\end{equation}
This last equation is equivalent to the projection condition, and no additional constraints are placed on $\eta$. We conclude that our supergravity solutions are in general $\f18-$BPS, as claimed in the main text. 

\section{Degeneration limits}    
\label{app:zeroes}          

In the main text we restricted ourselves to simple roots and double roots at $w=0$ and discussed the most relevant ingredients. In this appendix, we complete this picture and discuss the full parameter space of solutions with vanishing $Y_I$. The nature of such solutions is entirely determined by the degeneracy of the largest root $w_0$ of the function $f(w)$. We divide the discussion depending on the multiplicity of the largest root.

\textbf{Simple root: } Near a simple root we can approximate the metric as follows 
\begin{equation}
    \ds_\Sigma^2 \rightarrow \dd \rho^2 + \f{f'(w_0)^2}{w_0^2} \rho^2\,\dd z^2\,,
\end{equation}
where we defined $\rho^2 = \f{(w-w_0))}{f'(w_0)}$. The metric near the root approaches a $\bR^2/\bZ_k$ conical singularity, with $k= \f{|w_0|}{f'(w_0)\Delta\,z}$. In the limiting case when $w_0\to 0$ the metric becomes,
\begin{equation}
    \ds_\Sigma^2 = \dd \rho^2 + \f{\dd z^2}{\rho^2}\,,
\end{equation}
where $w= f'(0)r^2$. Near $\rho=0$ the radius of the circle and the Ricci scalar blow up. It is not clear if this solution has a physical origin and if so what kind of defect can be described by it. We will not consider it any further.

\textbf{Double root: } This case can be further divided in two cases, a double root at $w_0=0$ or at $w_0\neq 0$. In order to find a double root at $w=0$ we need to put two parameters to zero, e.g. $q_2=q_3=0$. In this case, the metric near the root becomes
\begin{equation}
    \ds_\Sigma^2 \rightarrow \dd \rho^2 + \f{q_1-1}{q_1} \dd z^2\,,
\end{equation}
where we defined $\rho = w/\sqrt{4(q_1-1)^3}$. Near the double root the metric becomes a metric on the cylinder. Note that the prefactor in the five-dimensional metric near this point vanishes so that the space still caps off.

The other option is to have a double root at $w_0\neq 0$. In this case the metric takes the form
\begin{equation}
    \ds_\Sigma^2 \rightarrow \f{1}{4f''(w_0)}\left(\dd \rho^2 + \f{f''(w_0)^2}{w_0^2}\e^{-2\rho} \dd z^2\right)\,,
\end{equation}
where we defined $w-w_0 = \f12\e^{-r}$. This is the metric on $\bH^2/\bZ_k$, where $k=\f{|w_0|}{\Delta \,z f''(w_0)}$. It is not clear what kind of defects (if any) are described by these geometries so we won't consider this case any further.

\textbf{Triple root: } As in the previous cases, we will distinguish the case when the triple root occurs at $w_0=0$ or $w_0\neq 0$. The former case occurs when $q_2=q_3=0$ and $q_1 = 1$. In this case the metric approaches
\begin{equation}
    \ds^2 = \f{1}{2\rho}\left(\dd\rho^2 + \dd z^2\right)\,,
\end{equation}
where we defined $\rho = \f1{2w}$. The full two-dimensional space, blows up near $\rho=0$. It is not clear what kind of defects (if any) are described by these geometries so we won't consider this case any further.

Finally, we can consider a triple root at $w=w_0 \neq 0$. In this case the metric reduces to
\begin{equation}
    \ds^2 = \f{1}{8\sqrt{w_0}\rho^{3/2}}\left(\dd\rho^2 + \dd z^2\right)\,,
\end{equation}
where $\rho=\f{w_0}{4(w-w_0)^2}$. Again we find a metric that blows up near $\rho=0$ for which the physical interpretation (if any) remains obscured.

Given this analysis we will restrict ourselves to the situations where the biggest root is a simple root or a double root at $w_0=0$. As discussed above, the latter situation occurs when two of the $q_I$'s vanish and the third is larger than one. The case with a simple root is the generic case, which occurs whenever the discriminant of $f$ is non-vanishing. Alternatively, the largest root can be a simple root when the discriminant is zero with the double root smaller then the simple. assuming the form of $f$ to be $f=(w-w_a)^2(w-w_b)$, this occurs in various scenarios,
\begin{align}
    w_b> w_a \quad \Leftrightarrow \quad \begin{cases}
        q_2 = q3 = 0 & \quad \text{and}\quad  q_1<1\,,\\
        q3 = 0\,, q2=1+q_1 \pm 2\sqrt{q_1} & \quad \text{and}\quad  q_1 \pm \sqrt{q_1} > 0 \,,\\
        \Delta(f) = 0 & \quad \text{and}\quad  q_3 < q_3^*  \,,
    \end{cases}
\end{align}
where $q_3^* = \f12 \left(2+q_1+q_2-\sqrt{3}\sqrt{4q_1+4q_2-q_1^2-q_2^2+2q_1q_2}\right)$. This last constraint is hard to solve in all generality but for separate instances is easy to verify.

\section{Uplift to ten dimensions}    %
\label{app:uplift}           %

Every solution of maximal $\SO(6)$ gauged supergravity in five dimensions can be uplifted to type IIB supergravity \cite{Cvetic:1999xp,Cvetic:2000nc}. The uplifted solution has a constant dilaton $\e^\Phi = g_s$ and the following non-vanishing fields,
\begin{align}
    \ds_{10}^2 =&\, {\Delta}^{1/2} \ds_5^2 + g^{-2} \tilde{\Delta}^{-1/2} T_{ij}^{-1}\DD\mu^i\DD\mu^j\,,\\
    F_5 =&\, G_5 + \star G_5\,,\\
    G_5 =&\, -g U \epsilon_{(5)} + g^{-1}\left( T_{ij}^{-1} \star \DD T_{jk} \right) \wedge \left(\mu^k \DD\mu^i\right)- \f12 g^{-2} T_{ik}^{-1} T_{jl}^{-1} \star F_{(2)}^{ij}\wedge \DD\mu^k\wedge \DD\mu^l\,.
\end{align}
In the expressions above we defined 
\begin{equation}
\begin{aligned}
    U=&\, 2 T_{ij}T_{jk} \mu^i\mu^k- {\Delta} T_{ii}\,,\qquad & \tilde{\Delta}=& T_{ij}\mu^i\mu^j\,,\\
    F^{ij}_{(2)} =&\, \dd A^{ij}_{(1)} + g A_{(1)}\,,\qquad & \DD\mu^i =&\, \dd\mu^i + g A^{ij}_{(1)} \mu^j\,,\\
    \DD T_{ij} =&\, \dd T_{ij} + g A^{ik}_{(1)} T_{kj} + g A^{jk}_{(1)} T_{ik}\,. 
\end{aligned}
\end{equation}
The $\mu^i$ are embedding coordinates for the five-sphere such that $\mu^i\mu^i = 1$. In order to identify the fields with those in the main text we have
\begin{align}
    A^{(I)} =&\, A_{(1)}^{2I-1\,, 2I}\,,\\
    T_{ij} =&\, \diag \left( X_1\e^{Y_1}\,,X_1\e^{-Y_1}\,,X_2\e^{Y_2}\,,X_2\e^{-Y_2}\,,X_3\e^{Y_3}\,,X_3\e^{-Y_3} \right)\,.     
\end{align}
Note that here we have fixed the gauge in order for the off-diagonal fields to vanish. In general there would be additional off-diagonal terms of the form $T_{2I-1 \,, 2I} \sim \ii\arg Y_I$. 

\subsection{Fefferman-Graham coordinates}

In order to study the asymptotic behaviour of the uplifted solution it is convenient to work in Fefferman-Graham coordinates. We consider the uplifted metrics found in Section \ref{sec:uplift}, and use the constrained coordinates
\begin{equation}
    \mu_1 = \sin{\theta}, \quad \mu_2 = \cos{\theta} \sin{\psi}, \quad \mu_3 = \cos{\theta} \cos{\psi}
\end{equation}
We wish to perform the change of coordinates $(w,\theta,\psi) \rightarrow (\rho,\beta,\alpha)$. Finding the appropriate coordinate transformation amounts to solving the equation:
\begin{equation}
    f_{w}^{2} \dd w^{2} + f_{\theta}^{2} \dd \theta^{2} + f_{\psi}^{2} \dd \psi^{2} + f_{\theta \psi}^{2}  \dd \theta \dd \psi = \frac{\dd \rho^{2}}{4\rho^{2}} + \left( \hat{g}_{\beta} \dd \beta^{2} + \cos^{2}{\beta} \hat{g}_{\alpha} \dd \alpha^{2} + \hat{g}_{\alpha \beta} \dd \alpha \dd \beta \right) \, ,
\end{equation}
Solving order by order in the new coordinates yields the explicit transformation:
\begin{equation}
\begin{split}
    w \rightarrow & \, \frac{1}{\rho} + \frac{1}{4} [2+ (q_{1}+q_{2}\cos^{2}\alpha +q_{3}\sin^{2}\alpha)\cos ^2\beta-2 (q_{1}+q_{2}+q_{3})\\&+ (q_{2}+q_{3}) \sin ^2\beta ]+c_{1}(\alpha,\beta) \rho + c_{2}(\alpha,\beta)\rho^{2} \dotsm \\
    \theta \rightarrow & \, \beta + \frac{1}{16} \sin (2 \beta ) (-q_{1}+q_{2}\sin^{2}\alpha +q_{3}\cos^{2}\alpha ) + b_{1}(\alpha,\beta) \rho + b_{2}(\alpha,\beta) \rho^{2} + \dotsm\\
    \psi \rightarrow & \, \alpha + \frac{1}{16} \sin (2\alpha) (q_{3}-q_{2}) + a_{1}(\alpha,\beta) \rho + a_{2}(\alpha,\beta) \rho^{2} + \dotsm
\end{split}
\end{equation}
The coefficients are now angular functions of $(\alpha,\beta)$, and were determined up to quadratic order in $\rho$, although their expressions are omitted as they are very cumbersome. We used these coordinate transformations in the main text to determine the form of the cutoff $\Lambda(\epsilon_{\rho},\theta,\psi)$ in Eq. \eqref{eq:lambda_cutoff}. 

For completeness, we also report the form (to first order) of the metric functions for the angular coordinates. These are:
\begin{equation}
\begin{split}
    \hat{g}_{\alpha} = 1 -& \, \frac{\rho}{32} (12 q_{1} \cos (2 \beta )+4 q_{1}+3 q_{2} \cos (2 (\alpha +\beta ))-6 q_{2} \cos (2 \beta )\\&+3 (q_{2}-q_{3}) \cos (2 (\alpha -\beta ))-6 \cos (2 \alpha ) (q_{2}-q_{3})\\&-2 q_{2}-3 q_{3} \cos (2 (\alpha +\beta ))-6 q_{3} \cos (2 \beta )-2 q_{3})+ \dotsm \\
    \hat{g}_{\beta} = 1 + & \, \frac{\rho}{8} (2 q_{1}-3 \cos (2 \alpha ) (q_{2}-q_{3})-q_{2}-q_{3}) + \dotsm\\
    \hat{g}_{\alpha \beta} = \frac{3}{8}& \, \rho  \sin (2 \alpha ) \sin (2 \beta ) (q_{3}-q_{2})+ \dotsm
\end{split}
\end{equation}
In these coordinates it is also easy to check that in the limit $\rho \rightarrow 0$ we retrieve the metric of $AdS_{5} \times S^{5}$, as expected.

The discussion is analogous when $Y_1 \neq 0$. We have found an uplifted solution only in the $\f12-$BPS case, with $q_2 = q_3 = 0$. In this case we find
\begin{equation}
\begin{split}
    w \rightarrow & \, \frac{1}{\rho} + \frac{1}{4} [2+ q_1 \cos ^2\beta -2 q_{1} + K \sin^2\beta \cos 2\phi_1]+c_{1}(\beta) \rho + \dotsm
\end{split}
\end{equation}
Which determines the form of the cutoff in Eq. \eqref{eq:lambda_cutoffY}.

\newpage
\bibliographystyle{JHEP}
\bibliography{theBib}		
\end{document}